\documentclass{article}
\usepackage{graphicx} 
\usepackage[legalpaper, margin = 1.3cm]{geometry}
\usepackage{amssymb}
\usepackage{amsmath}
\usepackage{xcolor}
\usepackage{bm,url,wasysym}
\usepackage[square,sort,comma,numbers]{natbib}
\usepackage{authblk}

\title{Probing beyond-$\Lambda$CDM cosmology with Gravitational Waves}

\author{Kabir Chakravarti}

\affil[1]{CEICO, FZU-Institute of Physics of the Czech Academy of Sciences,
Na Slovance 1999/2, 182 21 Prague 8, Czech Republic}

\begin{document}

\maketitle

\begin{abstract}
The propagation of Gravitational Waves has been reliably recognised as a test-bed for beyond standard models of gravity and cosmology. We utilise this property to examine the effects of a class of parametrised beyond-$\Lambda$CDM cosmology on inferece of GW parameters. We find that the combined beyond-$\Lambda$CDM likelihood function exhibits correlations between the parameters which are especially dependent upon binary eccentricity. Expanding on previous results, we demonstrate through Fisher forecasts that we would need nearly 1 year of 3G GW data to be able to infer the beyond-$\Lambda$CDM model to $2\sigma$ significance. We also find counter-intuitively that errors of source-modelling leave large biases upon the inference of the beyond-$\Lambda$CDM parameters which come into play only during GW propagation.     
\end{abstract}

\section{Introduction} Cosmology with Cold Dark Matter and Dark Energy, known commonly as the $\Lambda$CDM model is quite robust as a model, and is able to explain most observations to date with reasonable success. A few mentionable examples are the temperature and polarisation spectra of Cosmic Microwave Background (CMB) and Baryon Acoustic Oscillations as observed by the WMAP \cite{WMAP:2008ydk} and Planck \cite{Planck:2018vyg} surveys. However as the data have improved, various tensions have also started to appear between data sets and continue to get stronger with newer data. The $H_0$ tension \cite{DiValentino:2021izs} and the $S_8$ tension \cite{Abdalla:2022yfr} are noteworthy examples. Such tensions are possibility related to the period of dark-energy domination at low redshifts. Other works have shown that the shortcomings of $\Lambda$CDM also involve violations to the cosmological assumptions of isotropy \cite{Migkas:2020fza, Secrest:2020has, Javanmardi:2015sfa} as well as homogeneity \cite{Gott:2003pf, Horvath:2013kwa} . These inadequacies make $\Lambda$CDM incomplete at the theoretical level. Motivated by these factors a host of so called beyond-$\Lambda$CDM theories have been proposed in order to mitigate all or some of the shortcomings of $\Lambda$CDM. Of them the Horndeski \cite{Clifton:2011jh} class of theories deserve special mention, as they are the most general class of scalar-tensor theories that are second order in their time derivatives, and therefore do not suffer from Ostrogradski instabilities in their solutions. When probes of geometry alone are considered, such as supernova distances and baryon acoustic oscilations, there is not much space for a significant deviation from LCDM background expansion history. Nonetheless, the cosmic tensions imply that a modification to $\Lambda$CDM structure formation is necessary In a seminal work, Bellini and Sawicki \cite{Bellini:2014fua} demonstrated that the space of scalar perturbations break down into four sub-spaces. Modifications to linear scalar perturbations are then described by four effective operators whose strength is determined by the set of parameters $\left\lbrace\alpha_K,\alpha_B,\alpha_M,\alpha_T\right\rbrace$. Of these four parameters, two, namely $\alpha_T$ and $\alpha_M$ also, separately, affect the propagation of gravitational waves.

\paragraph*{} It is in this context that sources of GWs become relevant. The observation of GWs from merging Black Holes (BHs) and Neutron Stars (NSs) \cite{LIGOScientific:2016aoc, LIGOScientific:2017vwq, LIGOScientific:2021qlt}  have ushered in a new era in astronomy and astrophysics. The sensitivity  of the network of current dectors, namely the LIGO-Virgo together with the newly joined KAGRA now allows us to detect events from as far away as a few thousands of Mpc. This implies that as GWs travel the intervening distance, their amplitude and phase can get modulated by the beyond-$\Lambda$CDM nature of the spacetime, and can be expected to show up as deviations during observations. This unique coincidence opens up a few interesting possibilities. To begin with, assuming a beyond-$\Lambda$CDM model one expects to understand the forecasts of its beyond-$\Lambda$CDM parameters resulting from a single observation. With a merger event detection rate of 10 per year on the average with the existing LIGO-Virgo-KAGRA network, we can expect a total detection tally to stand $\thicksim 100-200$ in a few years. Additionally, with the newer 3G network of ground-based detectors, namely the Einstein telescope and Cosmic Explorer expected to come online around 2030 the event detection rate would go up to tens of thousands a year.
In such a scenario, one can also consider the possibility of population-wide inference of the beyond-$\Lambda$CDM parameters. 

\paragraph*{} The observations from merging NSs in GW170817 have successfully demonstrated the utility of GWs in constraining the $\alpha_T$ subspace. It was reported \cite{Sakstein:2017xjx} that the difference in speeds of the graviton and the photon were constrained to better than 1 part in $10^{15}$. Constraints to $\alpha_T$ have also been obtained at similar magnitudes for very high energetic gravitons by considering the so-called `Gravito-Cherenkov' effect on highly energetic cosmic rays \cite{Moore:2001bv}. While being strong constraints, these observations nevertheless leaves room for interesting possibilities which form the basis considerations of works in this field. Previously, there were a few noteworthy studies along these lines, starting with \cite{Amendola:2017ovw, Belgacem:2018lbp} were one of the first to compute the impact of $\alpha_M$, the so-called `run-in-Planck-Mass' on GWs. Their results were then expanded upon by \cite{Finke:2021aom}, who considered the effect of the nature of populations of sources upon the inference of $\alpha_M$. These inferences relating to $\alpha_M$ were furthered by \cite{Mancarella:2021ecn, Leyde:2022fsc}. On the other hand, Baker et.al. as part of the cosmological working group of LISA \cite{LISACosmologyWorkingGroup:2022wjo} have come up with an in-depth analysis for the inference of the $\alpha_T$ subspace.

\paragraph*{} As we have seen, the $\alpha_M$ and $\alpha_T$ subspaces are the only subspaces to be probed by propagating GWs. It is therefore naturally logical to combine the results from the $\alpha_M$ and $\alpha_T$ subspaces into one combined inference, as opposed to standalone $\alpha_M$ or $\alpha_T$ inferences. This is basically the problem that we tackle in this paper. Specifically, we want to understand and provide an answer to the following questions

\begin{enumerate}
\item What are the forecasts of error for a combined $\alpha_M$ and $\alpha_T$ from a single merger event and what is the degree of covariance between the subspaces ? Further, what factors do these covariances depend on ?
\item Do effects at source (source parameters, or source modelling accuracy) affect the forecasts of propagation parameters ?
\item What size of a population-wide survey can lead to a meaningful inference of $\alpha_M$ and $\alpha_T$, and can it be done by the current detector network ? Additionally, will signatures specific to the type of population show up in the inference results ?
 
\end{enumerate}The remainder of the paper is organised as follows, in \ref{Sec:cosmo_prop}, we briefly review the preliminaries of cosmological propagation, and understand the formulation of how $\alpha_T$ and $\alpha_M$ affect the GW amplitude and the phase. Then in \ref{sec:src_mod}, we describe how $\alpha_T$ and $\alpha_M$ explicitly interact with the source via the Post-Newtonian framework. In \ref{sec:rslts}, we discuss our results. Our results are subdivided into three sections. Single events Fisher estimates are carried out in \ref{ssec:res_se}, which are followed up by single event PN studies in \ref{ssec:PN_acc}. Finally population studies are done in \ref{ssec:pop}.

\section{Cosmological preliminaries}\label{Sec:cosmo_prop}

Throughout this work, we assume the $\Lambda$CDM backgorund expansion history with Dark energy content $\Omega_\Lambda^0 = 0.689$, matter content $\Omega_m^0 = 1 - \Omega_\Lambda^0$ and $\mathcal{H}_0 = 70 $ km/sec/Mpc based on Plack data. Our model at background is exactly same to $\Lambda$CDM. As we mentioned before, to constrain such classes of theories identical to $\Lambda$CDM at the background one turns to perturbations. Our setup can be considered by the following, we assume GW arising out of compact binary merger events occuring in a redshift range $ z\leq 0.5$. In the course of travelling the intervening distance, the GWs will pick up the signatures of the beyond-$\Lambda$CDM cosmology, namely $\alpha_M$ and $\alpha_T$. The redshift range is  chosen, so that the events do not fall out of the sensitive volume of current generation detectors on one hand, while being sufficiently far away to accumulate enough effects of the non-$\Lambda$CDM signatures on the other. As we are interested in beyond-$\Lambda$CDM in the low curvature regime, we assume that the mechanics of generation of GWs is completely governed by General Relativity (GR). In the following sections, we focus exclusively on the subspace of interest $\alpha_M,\alpha_T$ and discuss the nature of the changes to GWs brought on by these effects.

\subsection{Effect of $\alpha_M$}\label{alphaM_effect}
Physically, $\alpha_M$ represents a a variation in the value of Planck mass with time. A variation in Planck mass shows up as a friction term in the perturbation equations. Our starting point is the modified GW propagation equation in presence of a variable Planck Mass according to \cite{Amendola:2017ovw} reads as

\begin{equation}\label{h_cosmo_evolution}
    \ddot{h} + \left[ 2 + \alpha_M(t) \right] \mathcal{H}(t) + \omega^2 h = \Gamma(t), 
\end{equation} with $\omega^2$  the frequency of the mode in question. $\alpha_M$ is in the most general case, a time-dependent function parametrising the modification in the friction. The source $\Gamma(t)$ on the right side operates as soon as there is anisotropic stress, or there exists a dark graviton coupled to the known massless graviton of GR. In this work, we will set $\Gamma = 0$ since we will be solely interested in the modification of the friction term in \ref{h_cosmo_evolution}. If we realise that \ref{h_cosmo_evolution} is of the general form

\begin{equation}
y^{''} + p(x)y^{'} + q(x)y = 0 \nonumber
\end{equation} then we may decompose the function $h$ into its constituents

\begin{equation}
h(T)=u(T) \cdot v(T), \; \; \text{with} \;  \; u(T) = \mathrm{exp}\left[-\frac{1}{2}\int dT \left (2+\alpha_m\right)\mathcal{H}(T)\right] \nonumber
\end{equation} and the function $v(t)$ is then easily found to obey the differential equation

\begin{equation}
     \ddot{v}(T) - f(T)v(T) = 0 \nonumber 
\end{equation} with 

\begin{equation} \label{eq:f(T)}
f(T)  \equiv \omega^2 \left[\left(1 + \frac{\alpha_m}{2}\right)^2 \left(\frac{\mathcal{H}}{\omega}\right)^2 + \left(1 + \frac{\alpha_m}{2}\right) \frac{\partial_T\mathcal{H}}{\omega^2} - 1 \right]
\end{equation}For systems such as astrophysical binaries $\mathcal{H} << \omega, \dot{\mathcal{H}} << \omega^2 $ , so that \ref{eq:f(T)} just boils down to $f(T) \approx -\omega^2 = $ constant, and then we must have 
 
\begin{equation} \label{fullsoln2}
    h(T) \propto \mathrm{exp}\left[-\frac{1}{2}\int dT \left (2+\alpha_m(t) \right)\mathcal{H}(t)\right] \times \mathrm{exp}[i \omega T].
\end{equation} This is a consequence of GWs being generated on super sub horizon lengthscales. Therefore, within this deep sub-horizon approximation, the contribution of the running Planck Mass $\alpha_m$ is completely washed out from the phase of the GWs, and lies only in dampening the amplitude. The Planck mass running $\alpha_m$ is a model dependent quantity, but for low redshifts we can assume $\alpha_M \approx $ constant

\begin{align}\label{h_damped_osc}
    h(T) &\propto \mathrm{exp}\left[- \left(1+\frac{\alpha_m}{2}\right)\mathcal{H} T \right] \times e ^{i\omega T} \nonumber \\
    h(z) &= \left[1+z\right]^{-\left(1 + \frac{\alpha_M}{2}\right)} h(z=0)
\end{align}We can now see that if we set $\alpha_M = 0$, we recover the usual $1/(1+z)$ redshift damping of $\Lambda$CDM cosmology. 

\subsection{Effect of $\alpha_T$}\label{alphaT_effect}
The presence of tensor modes of perturbation violate equivalence, and thereby allow for sub-luminal propagating modes. In literature, this is known as the tensor excess speed, and is often written as $\alpha_T = c_T^2 - 1$. We model the effects based exactly on the outline provided in \cite{LISACosmologyWorkingGroup:2022wjo}, but nevertheless provide a brief outline for completeness. We start with the parameter $\Delta$ defined in 2.12-2.13 of the reference

\begin{align}\label{Delta_def}
\Delta &= 1 - \frac{(c_T)_\mathrm{obs}}{(c_T)_\mathrm{src}} \nonumber \\
& = 1 - (c_T)_\mathrm{obs} 
\end{align} The source mechanics being Einstein gravity forces gravitons to be always emitted with $c = 1$, which sets $(c_T)_\mathrm{src} = 1$. Intervals of time at the source and at observation are thus related by

\begin{equation}\label{del_t_rel}
    dt_{\mathrm{obs}} = \left(\frac{1+z}{1-\Delta}\right) dt_{\mathrm{src}}
\end{equation}meaning that the instantaneous frequency evolution at the source and at the observer are then related by 2.18 of \cite{LISACosmologyWorkingGroup:2022wjo}, which reads

\begin{figure}
\centering
\includegraphics[width=7.5cm, height=5cm]{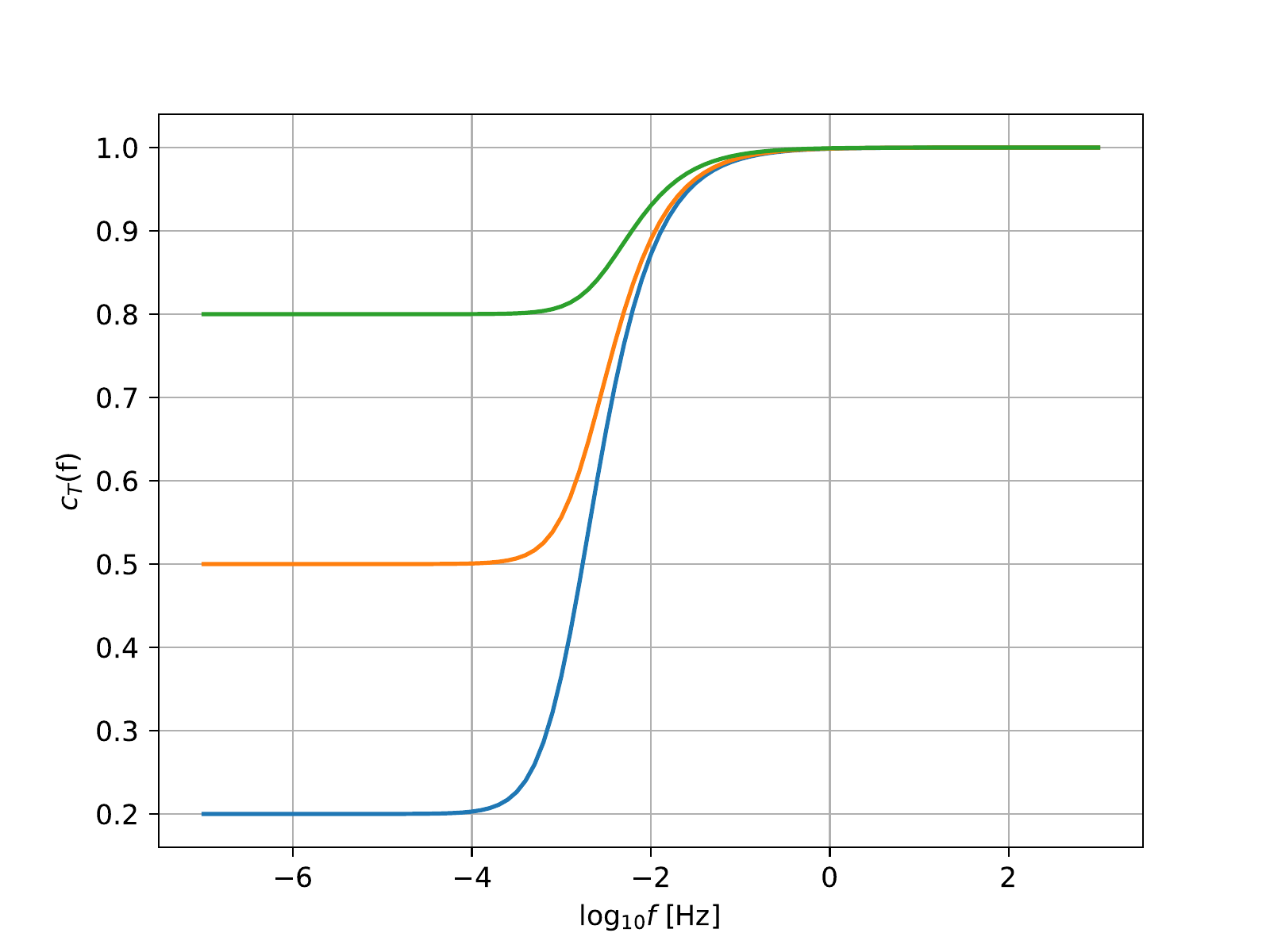}
\includegraphics[width=7.5cm, height=5cm]{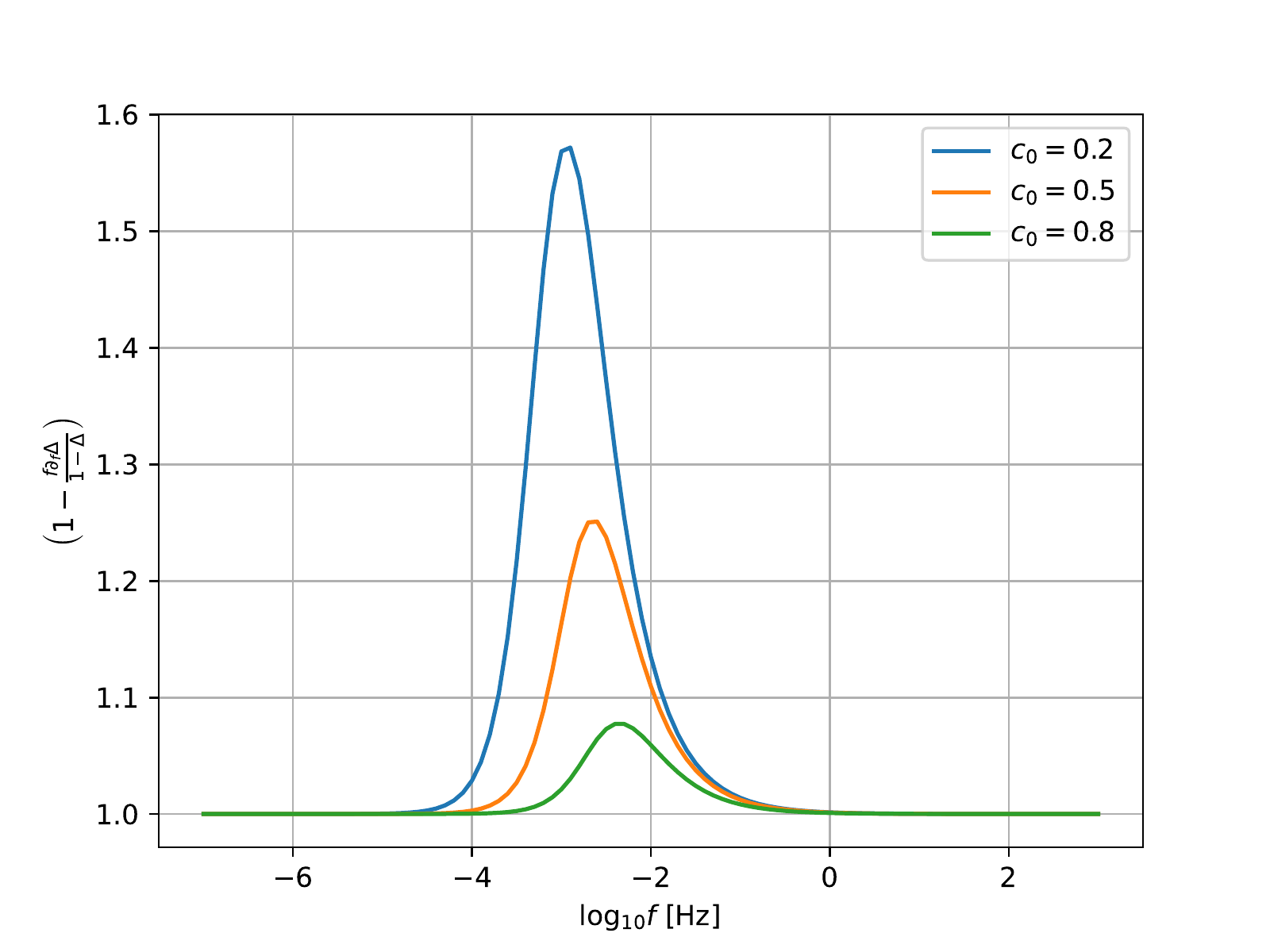}
\caption{$c_T$ (left panel) and the phase factor $\left(1 - \frac{\mathrm{d\quad log}(1-\Delta)}{ \mathrm{d\quad log}f}\right)$ for our models for different $c_0$}
\label{fig:ct}
\end{figure}

\begin{equation}\label{inst_f_evol}
\left(\frac{d\omega}{dt}\right)_\mathrm{src} = \left(\frac{1+z}{1-\Delta}\right)^2 \left(1 - \frac{\mathrm{d\quad log}(1-\Delta)}{ \mathrm{d\quad log}f_\mathrm{obs}}\right) \left(\frac{d\omega}{dt}\right)_\mathrm{obs}
\end{equation} Here the source value represents the true value, while the observed value is modulated by propagation effects. At this point we also note that following \ref{Delta_def} quantities get either redshifted (decrease) or blueshifted, so that certain combinations of quantities remain invariant under propagation. As we shall see in Section \ref{sec:src_mod} these propagation invariants are crucial in simplifying computations of GW amplitude and phase. To proceed further, we adopt $(c_T)_\mathrm{obs}$ to be 2.24 of \cite{LISACosmologyWorkingGroup:2022wjo}, the so-called EFT inspired ansatz. We reproduce the ansatz here for completeness.

\begin{equation}\label{EFT_ansatz}
c_T^2(f) = \left[1 + \left(\frac{f_*}{f}\right)^2 -  \left(\frac{f_*}{f}\right)^2 \sqrt{1 + 2(1-c_0) \left(\frac{f}{f_*}\right)^2} \right]
\end{equation}where f is the GW frequency at observation. Fig \ref{fig:ct} (left panel) shows the behaviour of $(c_T)_\mathrm{obs}$ as a function of frequency, with changing the parameter $c_0$ while keeping $f_*$ constant. It is clearly seen that $f_*$ is the frequency of transition from sub-luminal to luminal motion of the gravitons at the observer. . However, two counter arguments rescue us. First, the emission of EM counterparts relative to GW is somewhat model dependent, and hence any constraint derived is partucilar to a model of emission. The uncertainty of EM emission times across modelling partially weakens the otherwise extremely strong constraint. Secondly, and more importantly, it can also be possible that the transition frequeny $f_*$ is lower than the lowest frequency we could observe in a GW event with ground based detectors. In this case, we can still have sub-luminal gravitons and we would not even violate any bounds from GW170817 like events. Furthermore, as we mentioned before, the Gravito-Cerenkov constraints only affect very highly energetic gravitons ($10^{10}$ GeV) and are therefore not applicable to our scenario.

\paragraph*{}If \ref{inst_f_evol} is applied to orbital frequency, it is clearly seen that propagation non-trivially changes the phasing. In order to get the phasing at observation, the quantity of interest $[\Phi(t)]_\mathrm{obs}$ is computed by integrating the observed orbital frequency $[\Omega(t)]_\mathrm{obs}$. In Fig \ref{fig:ct} (right panel) we show the behaviour of the orbital frequency factor $\left(1 - \frac{\mathrm{d\quad log}(1-\Delta)}{ \mathrm{d\quad log}f}\right)$ where we use $c_T$ given by \ref{EFT_ansatz}  as a function of f with a range of chosen parameters for $c_0$ and $f_* = 0.1$ Hz, well below the lowest observed frequency bin for GW170817. To conclude the section, we note that $\alpha_T$ also affects the amplitude, in that it changes quantities like the perceived mass and luminosity distance at the point of observation, because of sub-luminal motion. It only remains to calculate each of the effects, which we will take up explicitly in the following section.

\section{Source modelling and explicit $\alpha_M,\alpha_T$ effects}\label{sec:src_mod}
It turns out that mergers of compact binaries emerge as possibly the best understood sources of GWs. A large part of this understanding is because of the existence of analytic or semi-analytic solutions spanning nearly the entire lifespan of the binary and also over a wide range of binary parameters. The solutions have historically been made possible because of the formulation of the PN framework, which aims to iteratively solve the field equations using the binary orbital velocity as an expansion parameter. The current state of the art sits at 4 PN order in orbital dynamics and corresponding GW phasing. It is remarkable that for comparable mass binaries, PN remains nearly consistent up to about $r \approx 6m$ or the Last Stable Circular Orbit (LSCO). Numerical simulations are needed only past LSCO to capture the merger.

\paragraph*{Phasing and $\alpha_T$}The idea behind binary phasing computation is simple. A binary with an average orbital separation $a$ continuously loses both energy $E$ and angular momentum $L$ due to emission of GWs. It turns out that both $E,J$ can be expanded in terms of the expansion parameter $x$
\begin{equation}\label{x_def}
    x = (m\Omega)^{2/3}
\end{equation}Here $\Omega$ is the angular velocity of the binary. Using Kepler's Law at leading order $\Omega^2a^3 = m$ we see that $x = v^2$ to leading order, where $v$ is the orbital velocity of the binary. Consequently the the energy flux $\mathcal{F}(x)$ and angular momentum fluxe $\mathcal{G}(x)$ relate as
\begin{align}\label{orb_evol1}
    \frac{dE(x)}{dt} &= -\mathcal{F}(x,e) \nonumber \\
    \frac{dL(x,e)}{dt} &= -\mathcal{G}(x,e) \nonumber \\
    m \frac{d\Phi_\mathrm{orb}(t)}{dt} &= x ^{3/2}
\end{align}where $e$ is the eccentricity of the binary in question. For circular binaries of comparable mass the relevant expressions for $E(x)$ and $\mathcal{F}(x)$ are computed to 3.5 PN order in \cite{Blanchet:2013haa}.\cite{Arun:2009mc} computes the expressions of $E(x),L(x,e)$ and their corresponding flux losses for the case of elliptic binaries of comparable mass. Finally, \cite{Barack:2003fp} forms our basis for the EMRI systems, where the equations are computed to 1 PN order in dissipation.  One then substitutes explicit expressions of $E(x),L(x,e), \mathcal{F}(x,e), \mathcal{G}(x,e)$ so obtained in \ref{orb_evol1} to obtain the differential equation for the simultaneous evolution of the orbital velocity $\Omega(t)$ and eccentricity $e(t)$.

\begin{align}\label{orb_evol2}
   \dot{\Omega}(t) &= \mathcal{M}^{5/3} \Omega^ {11/3} \quad \sum_i O_i(\eta, e) x^i \nonumber\\
   \dot{e}(t) &= -\mathcal{M}^{5/3} \Omega^ {8/3}e \quad \sum_i \mathcal{E}(\eta, e) x^i \nonumber\\   
   \Phi_\mathrm{orb}(t) &= \int dt\quad \Omega(t) 
\end{align}where $\eta$ is the symmetric mass ratio of the binary in question. The explicit expressions for the terms $\mathcal{O}(\eta, e)$ and $\mathcal{E}(\eta, e)$ have been taken from \cite{Arun:2009mc}.  Integrating \ref{orb_evol2} gives us the desired orbital phasing $\Phi_\mathrm{orb}(t)$. The GW phasing is just twice the orbital phasing, so $\Phi(t) = 2 \times\Phi_\mathrm{orb}(t)$. Considering the energy and angular momentum, and their corresponding fluxes to different $x$ (PN) powers, we end up with the corresponding ordered PN solutions for GTR. Incorporating the $\alpha_T$ effect now follows by substituting \ref{inst_f_evol} and \ref{del_t_rel} into \ref{orb_evol2}, so that we finally obtain 

\begin{align}\label{orb_evol_alphaT}
   \dot{\Omega}(t)_\mathrm{obs} &= \mathcal{M}_\mathrm{obs}^{5/3} \Omega_\mathrm{obs}^ {11/3} \quad \frac {\sum_i O_i(\eta, e) x^i}{\left(1 - \frac{\mathrm{d\quad log}(1-\Delta)}{ \mathrm{d\quad log}\Omega\mathrm{obs}}\right)} \nonumber\\
   \dot{e}(t) &= -\mathcal{M}_\mathrm{obs}^{5/3} \Omega_\mathrm{obs}^ {8/3}e \quad \sum_i \mathcal{E}(\eta, e) x^i \nonumber\\   
   \Phi_\mathrm{orb,obs}(t) &= \int dt\quad \Omega_\mathrm{obs}(t) 
\end{align}where $\Phi_\mathrm{obs}(t) = 2 \times\Phi_\mathrm{orb,obs}(t)$ Several remarks are in order. First, we note from \ref{x_def} that the variable $x$ is a propagation invariant. Physically, this just means that distance and time are stretched the same amount by expansion, so velocity $v \propto \sqrt{x}  $ has to be invariant. This is immensely useful at a computational level as it implies that propagation effects have not to be incorporated order by order, but rather as a multiplicative factor to the instantaneous angular velocity. Second, we see that $\alpha_T$ indeed explicitly modulates the phase by a multiplicative phasing factor, which is plotted in the right panel of Fig \ref{fig:ct}. It is clearly seen that it does tend to have a small $\mathcal{O}(1)$ but non-negligible effect. Finally, we see that the modulating factor does not appear in the eccentricity evolution equation explicitly. However as $e(t)$ is coupled to $\Omega(t)$ the observed eccentricity is also dependent on $\alpha_T$.

\paragraph*{Amplitude and $\alpha_T,\alpha_M$}To compute the amplitude, we note that it is only the quadrupolar amplitude that will survive asymptotically. They are computed by calculating the double time derivative of the source quadrupolar moments $M_{ij}$. ( See Eq 4.65 of \cite{10.1093/acprof:oso/9780198570745.001.0001} for a detailed expression). The corresponding GW polarisations for a line of sight oriented optimal binary in GTR thus becomes

\begin{align}\label{GWpol1}
    h_+(t) &= - 2 \left(\frac{\mu}{d}\right) [m\Omega(t)]^{2/3} \quad \left(\frac{2 \cos\left(2\Phi(t)\right) + e(t)\cos\Phi(t) \left(1 + 2\cos^2\Phi(t)\right) + e(t)^2}{1-e(t)^2}\right) \nonumber \\
    h_+(t) &= - 2 \left(\frac{\mu}{d}\right) [m\Omega(t)]^{2/3} \quad 2 \left(\frac{ \sin\left(2\Phi(t)\right) + e(t)\sin\Phi(t) \left(1 + 2\cos^2\Phi(t)\right) }{1-e(t)^2}\right) \nonumber \\           
\end{align}where $m, \mu$ are the source total and reduced mass respectively and $d$ is the physical distance to the binary. Under cosmological expansion in $\Lambda$CDM, we have the observed reduced mass $\mu_\mathrm{obs} = \mu/(1+z)$, where z is the redshift. Equivalently, for a given comoving distance $d_\mathrm{C} = d$, the proper distance scales as $d_\mathrm{P} = d_\mathrm{C}/(1+z) = d/(1+z)$. Finally, we also need to take into account that \ref{GWpol1} is expressed in source frame time, and thus in the observer's frame we pick up an additional factor of $1/(1+z)$. Putting all factors together, we see that the combination 

\begin{equation}
\left(\frac{\mu}{d}\right) \rightarrow \left(\frac{\mu}{d}\right) \left(\frac{1}{1+z}\right) \nonumber
\end{equation}in \ref{GWpol1}. Indeed, this is just the familiar amplitude damping factor. As we saw in \ref{h_damped_osc}, including $\alpha_M$ is achieved by multiplying by a factor of $(1+z)^{-\alpha_M/2}$. The effects of $\alpha_T$ can now be included. As evident from 3.7 and 3.1 of \cite{LISACosmologyWorkingGroup:2022wjo}, $\mu$ picks a factor of $c_T$, while $d$ contributes $1/\sqrt{c_T}$. So at last, the expression for the GW at the detector, including both $\alpha_M,\alpha_T$ becomes

\begin{align}\label{GWpol2}
    h_+(t) &= - 2 \left(\frac{\mu}{d}\right) \frac{c_T[2\Omega(t)_\mathrm{obs}]^{3/2}}{(1+z)^{(1 + \alpha_M/2)}} \quad [m\Omega(t)]^{2/3} \quad \left(\frac{2 \cos\left(2\Phi(t)\right) + e(t)\cos\Phi(t) \left(1 + 2\cos^2\Phi(t)\right) + e(t)^2}{1-e(t)^2}\right) \nonumber \\
    h_\times(t) &= - 2 \left(\frac{\mu}{d}\right)\frac{c_T[2\Omega(t)_\mathrm{obs}]^{3/2}}{(1+z)^{(1 + \alpha_M/2)}} \quad [m\Omega(t)]^{2/3} \quad 2 \left(\frac{ \sin\left(2\Phi(t)\right) + e(t)\sin\Phi(t) \left(1 + 2\cos^2\Phi(t)\right) }{1-e(t)^2}\right) \nonumber \\           
\end{align}

\section{Results}\label{sec:rslts}

We have performed three separate but related tasks, with GWs from binary inspirals over a wide variation of source characteristics. First, we made use of \ref{GWpol2} to calculate Fisher forecasts for single events. Here we have computed the $\alpha_T,\alpha_M$ forecasts from both circular and elliptic configurations of comparable mass binaries as well as EMRIs as opposed to \cite{LISACosmologyWorkingGroup:2022wjo} who make use of inspiral of circular EMRI systems only.  For our systems of comparable mass binaries, we have considered the phase evolution equations to 2 PN order beyond quadrupolar, following \cite{Arun:2009mc, Blanchet:2013haa} for both the circular and elliptic cases. Second, we have made use of the PN formalism to reduce accuracy order-by-order in \ref{GWpol2}, and study the corresponding effect upon the $\alpha_T,\alpha_M$ forecasts. Finally, we have also considered an exercise of population-wide inference of $\alpha_T,\alpha_M$ in order to get an idea of the volume of data necessary to adequately constrain the subspace. We primarily premise our work upon the inference obtained from ground-based detectors, namely the LVK network and the upcoming 3G detector network. As EMRIs are not relevant to ground based detectors, they are not our main focus and we have only included 1PN beyond-quadrupolar effects for them, following \cite{Barack:2003fp}. We have also limited our population-wide studies to comparable mass inspirals only because current characterisations of EMRI populations turn out to be heavily dependent on numerical N-body modelling of galactic nuclei environments \cite{Amaro-Seoane:2007osp}. In all our studies, our analysis is purely considering non-spinning binaries only.

It is evident from its formulation that the gravitational waveform $h(\vec{\theta})$ is a multivariate function, where $\vec{\theta} = \left\lbrace M, \eta, \alpha_m, c_0, f_* \right\rbrace$. The functional dependencies separate out into dependencies at the source in $M, \eta$ and dependencies during propagation in $\alpha_m, c_0 \& f_*$. We ultimately want to run a simultaneous Bayesian inference upon both the source and propagation parameters. However as Bayesian MCMC is computationally expensive, one normally performs a computationally cheap Fisher error estimate. The Fisher estimates are obtained by sampling near the peak of the likelihood function. As is well known \cite{Vallisneri:2007ev}, this means that the Fisher results are a good approximation only in the Linear Signal Approximation (LSA).  
   
\begin{equation}\label{LSA}
h(\vec{\theta}) = h(\vec{\theta}_0) + \partial_i h \Delta\theta^i 
\end{equation}. Now with \ref{LSA}, it is evident that 

\begin{equation}\label{lklhd_LSAb}
p(d|\vec{\theta}) \propto \mathrm{exp}\left[ -\frac{1}{2}\frac{\langle \partial_jh|\partial_ih\rangle}{\langle h|h \rangle^2} \Delta\theta^i\Delta\theta^j \right]
\end{equation} Evaluating the covariance from the distribution, one can see that it is proportional to the inverse of the Fisher Matrix $\langle \partial_i h|\partial_jh\rangle^{-1}$.  However, for real signals away from LSA the inverse of the Fisher matrix can only be regarded as a lower bound upon the error covariance matrix. This is also well known as the Cramer-Rao bound.

\subsection{Single event estimates}\label{ssec:res_se}

\begin{table}[h] 
\begin{tabular}{ |p{0.8cm}|p{0.8cm}| p{0.5cm}||p{0.5cm}|p{0.5cm}|p{0.6cm}||p{1cm}|p{0.8cm}|p{0.8cm}||p{1.1cm}|p{1.1cm}|p{0.8cm}|p{0.8cm}|p{0.8cm}|p{0.8cm}||p{0.5cm}|  }
 \hline
 \multicolumn{3}{|c|}{Source params} & \multicolumn{3}{|c|}{Cosmo params} & \multicolumn{3}{|c|}{Noise} & \multicolumn{6}{|c|}{Errors} & \multicolumn{1}{|c|}{}\\
 \hline
Chirp Mass $\mathcal{M}$ & Red. mass $\mu$  &  $e_i$ & $\alpha_m$ & $c_0$ & $f_*$ & PSD & lower cutoff & upper cutoff & $\delta(\mathrm{ln}\mathcal{M})$ & $\delta(\mathrm{ln}\mu)$ & $\frac{\delta\alpha_m}{\alpha_m}$ & $\frac{\delta c_0}{c_0}$ & $\frac{\delta f_*}{f_*}$ & S/N & 
 \\
 \hline
 10.0 & 5.0 & 0.00 & 0.7 & 0 & - & aLIGO & 20 & 200 & 0.001 & 0.004 & 3.018 & - & - & 9.93 & 2PN  
 \\ 
 \hline
 10.0 & 5.0 & 0.25 & 0.7 & 0 & - & aLIGO & 20 & 200 & 0.003 & 0.013 & 3.473 & - & - & 9.98 & 2PN  
 \\  
 \hline
 10.0 & 5.0 & 0.00 & 0.7 & 0.2 & 0.002  & aLIGO & 20 & 200 & 0.004 & 0.027 & 3.045 & 0.849 & 0.125  & 9.98 & 2PN 
 \\ 
 \hline
 10.0 & 5.0 & 0.25 & 0.7 & 0.2 & 0.002  & aLIGO & 20 & 200 & 0.006 & 0.045 & 4.296 & 0.397 & 0.049  & 10.08 & 2PN 
 \\ 
  \hline
 20.0 & 11.3 & 0.00 & 0.7 & 0.2 & 0.002  & aLIGO & 20 & 200 & 0.010 & 0.037 & 3.108 & 1.571 & 0.092  & 10.04 & 2PN 
 \\ 
  \hline
 20.0 & 11.3 & 0.25 & 0.7 & 0.2 & 0.002  & aLIGO & 20 & 200 & 0.012 & 0.035 & 3.208 & 0.495 & 0.179  & 9.97 & 2PN 
 \\  
 \hline 
 \hline
   2000.0 & 25.0 & 0.00 & 0.7 & 0 & - & LISA & $10^{-5}$ & 0.1 & $5.15 \times 10^{-7}$ & $2.76 \times 10^{-6}$ & 2.971  & - & - & 10.08 & 1PN 
 \\ 
 \hline
   2000.0 & 25.0 & 0.25 & 0.7 & 0 & - & LISA & $10^{-5}$ & 0.1 & $5.82 \times 10^{-7}$ & $7.66 \times 10^{-7}$ & 2.974  & - & - & 10.07 & 1PN 
 \\ 
 \hline 
   2000.0 & 25.0 & 0.00 & 0.7 & 0.2 & 0.002 & LISA & $10^{-5}$ & 0.1 & $1.80 \times 10^{-6}$ & $6.82 \times 10^{-6}$ & 3.000  & $1.48 \times 10^{-5}$ & $7.65 \times 10^{-6}$ & 9.99 & 1PN 
 \\ 
 \hline  
   2000.0 & 25.0 & 0.25 & 0.7 & 0.2 & 0.002 & LISA & $10^{-5}$ & 0.1 & $1.65 \times 10^{-6}$ & $7.12 \times 10^{-6}$ & 3.002  & $9.56 \times 10^{-5}$ & $8.68 \times 10^{-5}$ & 9.98 & 1PN 
 \\ 
 \hline   
\end{tabular}
\caption{Single event fisher estimates of the errors associated with different configurations of binary systems. Notice that the different systems have to be observed under different PSDs with separate range of sensitivities to give meaningful results. All the fisher estimates have been normalised to S/N $\approx 10$. \label{res_single_event}}
\end{table} 

\paragraph*{}\ref{res_single_event} shows the results of Fisher forecasts for a single GW event. To be consistent across systems we have considered the results at a signal to noise factor S/N of 10. Having EMRI and comparable mass results side by side helps us to study the comparative strengths of either system. We immediately notice that despite being truncated at 1PN order lower, the errors of the forecasts for EMRI systems for all the parameters except $\alpha_M$ are much less  compared to the systems of comparable mass. This improvement is because of improved accuracy arising from the relatively longer inspiral timescales in the EMRI system. The improvement is particularly pronounced for the estimates for $c_0, f_*$ where the errors associated with comparable mass systems are $\thicksim 10 \%$ but for EMRIs are $\thicksim 0.001 \%$ meaning that the errors decrease by a factor of $10^{-4}$. In addition to having longer inspiral, forecasts for an injected $f_* = 0.002 $ Hz are further improved as the transition from sub-luminal to luminal motion for gravitons occurs at a frequency which is within the sensitivity band of eLISA like detectors. In contrast, we see that the estimation of $\alpha_M$ is hardly affected by the system in question. Furthermore the error in $\alpha_M$ for every kind of system is $\thicksim 300 - 400 \%$, which is quite high compared to the other parameters. The inaccuracy itself should not be surprising, because $\alpha_M$ does not appear in the phase and also because it is degenerate with the redshift of the system. Additionally, both EMRI and comparable mass systems have the same functional dependences for their asymptotic amplitudes -- which would explain why estimation of $\alpha_M$ does not change between EMRI and comparable mass systems.

\begin{figure}
\includegraphics[width=20cm, height=12cm]{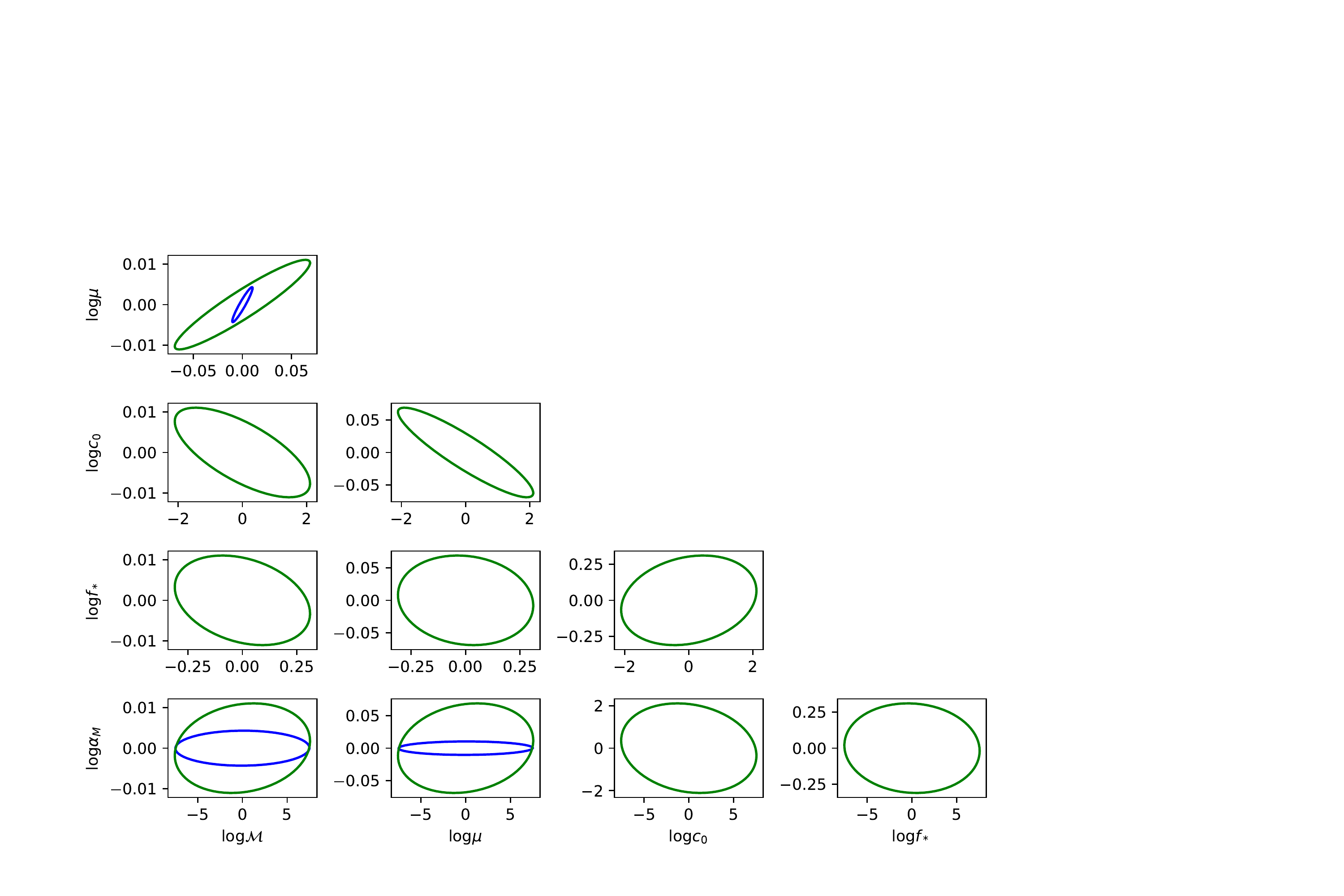}
\includegraphics[width=20cm, height=12cm]{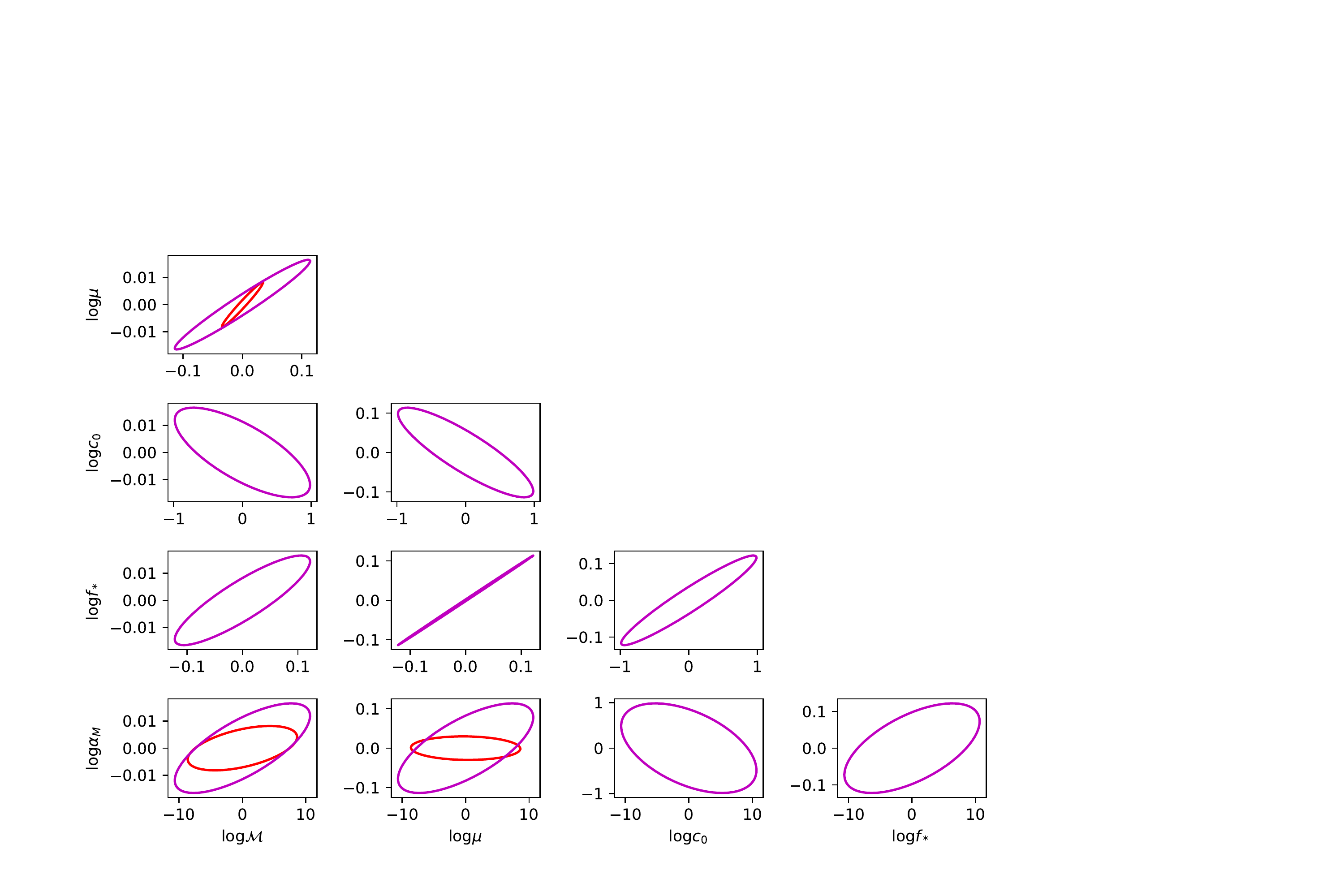}
\caption{Single event forecast results for comparable mass binary systems with $e=0$ (top) and $e=0.25$. Blue/Red plots represent simultaneous forecast of $\mathcal{M},\mu,\alpha_M$, while green/pink include $\alpha_T$ effects as well. In addition to affecting the variances marginally, the eccentricity does seem to affect covariances }
\label{fig:stst_ellipse}
\end{figure}

\paragraph*{} Fig \ref{fig:stst_ellipse} plots the Fisher ellipses of the comparable mass binary systems, for 0 and non 0 eccentricities. We note that the $e=0$ case shows mild covariances in the $(\alpha_M, f_*)$rows, as is evident from the horizontal nature of the ellipses. We see that for the $e=0.25$ case, the corresponding ellipses get tilted. This means that the likelihood function has non-trivial eccentricity dependences, which show up in the fisher plots as eccentricity dependent covariances. It is thus clear that eccentric binaries will behave as nuisances by introducing unnecessary bias in case our inference model is specific to circular inspirals. Indeed, this can be thought of a strong motivation for the accurate modelling of eccentric systems for both EMRI and comparable mass.

\subsection{Changing the PN accuracy}\label{ssec:PN_acc}
In \ref{ssec:res_se}, we demonstrated the forecasts of a single event for different binary configurations. So we might ask ourselves: how important really is it for us to model the source ? In other words, if we were to decrease the PN accuracy order by order, would it have any effect upon the estimates. Before proceeding, it is worthwhile to pause, and describe the nature of the PN terms order by order. As is evident from \ref{sec:src_mod}, \ref{GWpol2} is  obtained by considering the expression of the $\mathcal{F}(x,e)$ and $\mathcal{G}(x,e)$ upto a 2 powers in $x$ beyond the leading order. In order to do this, we have had to account for the contribution of the leading order hereditary (or tail) term appearing at order $x^{3/2}$ for both $\mathcal{F}(x,e)$ and $\mathcal{G}(x,e)$, which requires a careful consideration of the calculation  of the so-called eccentricity tail enhancement functions $\phi(e)$ and $\tilde{\phi}(e)$ respectively. Analytically, these functions are infinite series of Bessel functions with the eccentricity $e$ as their argument.  We have calculated these functions by fitting to their numerical values as presented in Appendix B of \cite{Arun:2009mc}. Fig \ref{fig:ecc_enhnce} shows the result of the fitting. The tiny residuals compared to the function values indicate a very good fit, with the maximum error being of order $0.1\%$.    

\begin{figure}[h]
\centering
\includegraphics[width=10cm, height=8cm]{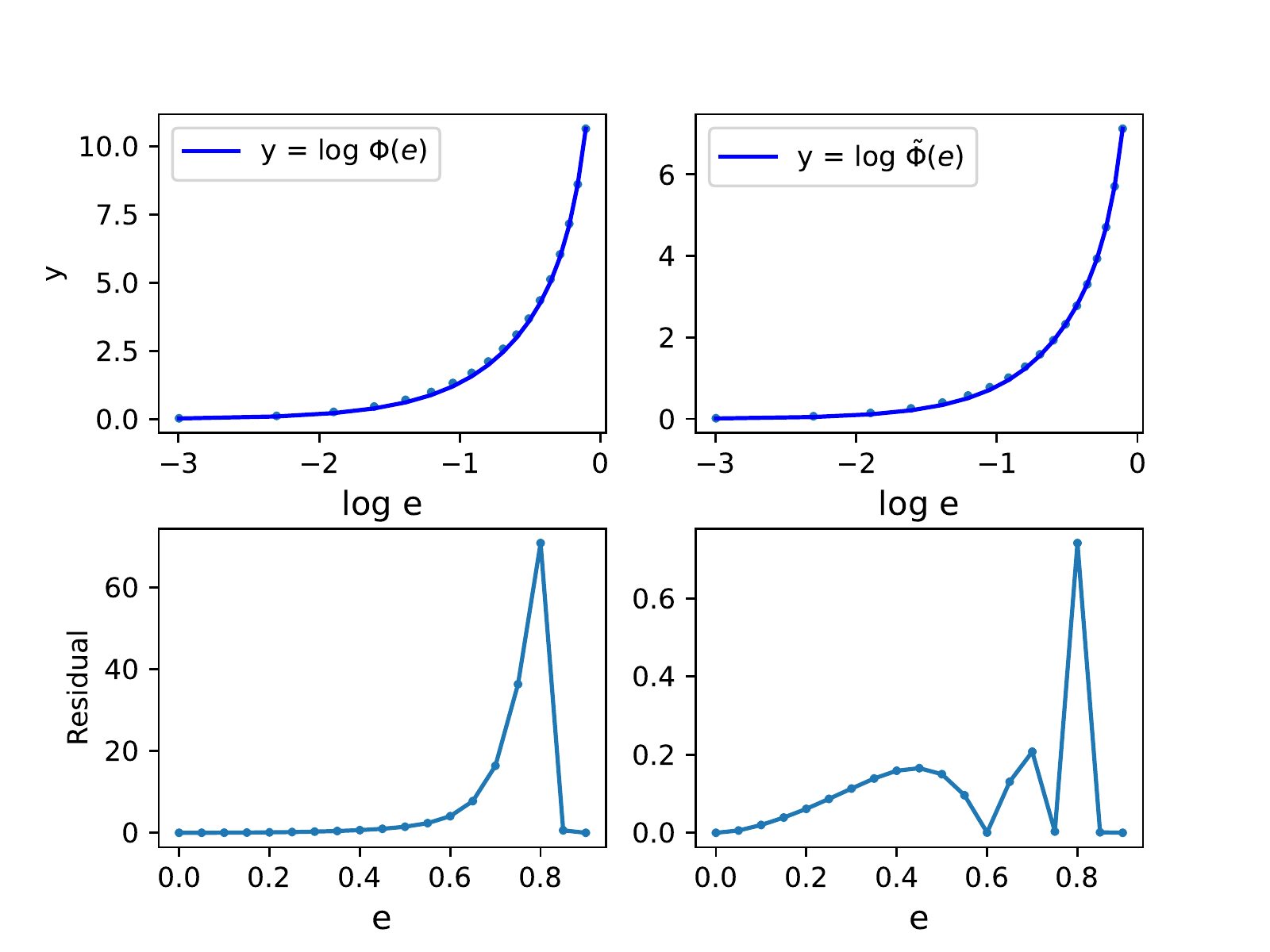}
\caption{Fitting for the eccentricity enhancement functions $\phi(e)$ and $\tilde{\phi}(e)$ for the tail contributions. The residuals are $\leq 0.3 \%$ for $\phi(e)$ and $\leq 0.2 \%$ for $\tilde{\phi}(e)$}
\label{fig:ecc_enhnce}
\end{figure}
\begin{figure}[h]
\centering
\includegraphics[width=8cm, height=6cm]{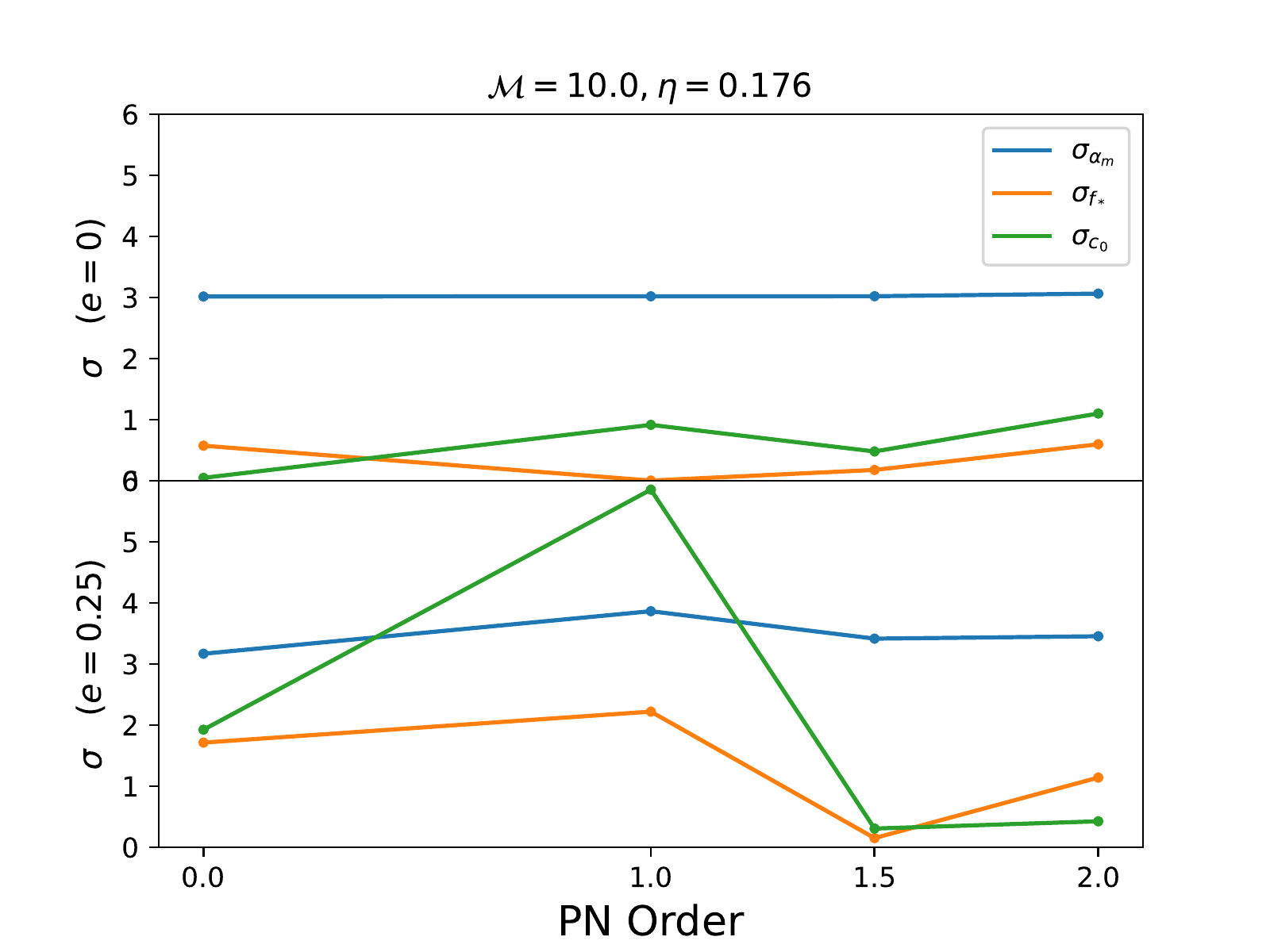}
\includegraphics[width=8cm, height=6cm]{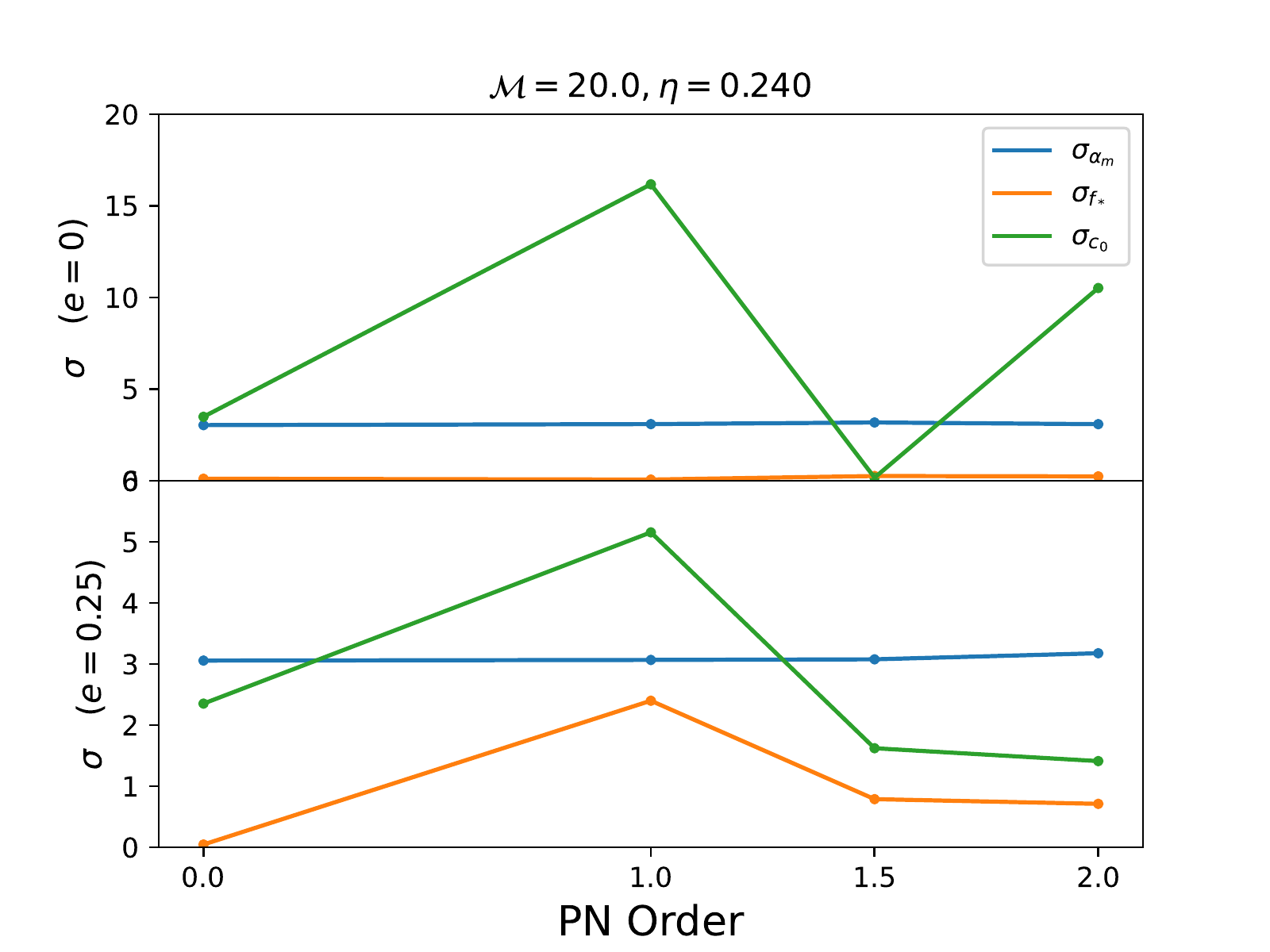}
\caption{Errors in Fisher forecasts in $\alpha_m, c_0, f_*$ as a function of the PN order at source, for two instances of comparable mass binary systems. Large variations are seen in the error estimates, notably for $c_0$ across the systems.}
\label{fig: PN_error}
\end{figure}

Let us now consider the case with comparable mass binaries. The results in \ref{res_single_event} have been obtained using the highest (2PN) order considered. However, what happens if we reduce the highest order ? To do this, we note that in addition to accounting for the expressions of $E,L,\mathcal{F,\mathcal{G}}$, we also need to account for the order-by-order correction to the Kepler's 3rd Law. In our work, we have used the PN-corrected Kepler's 3rd Law to calculate the corresponding separation $a$ for a given orbital frequency $\Omega$, obtained by integrating \ref{orb_evol_alphaT}. We terminate our evolution when the separation equals the Last Stable Circular Orbit (LSCO), i.e $a\leq r_{\mathrm{LSCO}}$. As is well understood, crossing the LSCO initiates a radial infall or plunge, and can be considered as a timestamp when the inspiral process terminates and merger begins. With this framework in place, we reduce the PN accuracy both from the analytic expressions, as well as from the Kepler's Laws. Fig \ref{fig: PN_error} shows the results of performing such an operation.

\paragraph*{} Before discussing the results we should note the factors that affect estimation accuracy, when PN order is played with. It is easy to see that the binary phasing is directly affected
as terms are added or subtracted to the explession of $E$ or $L$ and their corresponding fluxes. As PN ordering also plays with the Kepler's Law, they also affect the upper cut-off or termination frequency of the inspiral, thereby changing the length of the inspiral phase order-by-order, provided all orders start from the same starting frequency. As it involves a combination of factors, the estimation accuracy should not be expected to exhibit monotonic behaviour with changing PN prder. We can now focus on the results of the operation, as depicted in Fig \ref{fig: PN_error}. The figure on the left computes the order-by-order estimate for the $\mathcal{M} = 10, \mu = 5$ system, while the one on the right does the same for the $\mathcal{M} = 20, \mu = 11.3$ system. The upper panels represent circular configurations, while the lower ones represent the cases with initial eccentricity $e_0 = 0.25$. The PN-ordered accuracy computations reveals several interesting facts. To begin with, we note that the Fisher forecasts (or errors) for $ \alpha_T = (c_0, f_*)$ are more sensitive to PN corrections than the errors of $\alpha_M$. This is hardly surprising. PN terms alter the phase, not the amplitude. The amplitude is unaffected with $x$ because higher order moments do not survive asymptotically. This also explains why the $\alpha_M$ errors are unaffected even by changes in the PN order. Next, we note that for both circular and eccentric binaries the PN-ordered $\alpha_T$ errors are dependent on the configuration. As configurations directly play with binary phasing this effect should also be expected. Most importantly, we find that the estimates of $\alpha_T$ do vary non-trivially across the PN-order. We find that estimates are affected for both circular and eccentric binaries. While $\alpha_T = (c_0,f_*)$ estimates are affected in general, $c_0$ estimates are seen to be particularly extremely effected while playing with the PN order. This would possibly suggest that the change in PN order is capable of mimicking the kind of dephasing introduced by $\alpha_T$ through $c_0$ as shown in Fig \ref{fig:ct}. Furthermore, in all the cases, it can be observed that the estimates for $\alpha_T$ fluctuate at 1.5 PN order. As the tail terms are known to enter at 1.5 PN order, this means that such terms play a large role in the overall error budget. 

\paragraph*{} Due to relevant terms being known to  higher PN order, we chose comparable masses systems. Although performed for comparable mass binary systems, there are important lessons for the case of the EMRI systems as well. EMRI systems have much longer inspirals, so the relative effects of adding/subtracting terms order-by-order is expected to be greater. Second, this calculation indirectly  highlights the absolute importance of accurately the tail and tail-of-tail eccentricity terms for PN systems. This is especially true, because tiny discrepancies at the beginning will add up over much longer inspiral EMRI timescales -- and have the potential to give rise to huge errors in the measurement of $\alpha_T$

\paragraph*{} We are thus led to infer that modelling of source (considered here in PN ordered effects) does contribute non-negligibly towards errors of cosmological propagation if they appear in the phasing as seen by the observer. This exercise also demonstrates the importance of the leading tail contribution for the case of comparable mass systems.  Specifically, large and non-trivial errors can be expected in case of binaries where tail effects were not taken into consideration.

\subsection{Population studies}\label{ssec:pop}
The results obtained in \ref{ssec:res_se} demonstrate that although inferences from single merger events are quite a powerful tool to infer dynamical properties of the binary, they are not nearly enough for inference of the cosmological propagation parameters, namely $\alpha_T = (c_0,f_*)$ and $\alpha_M$, which are of $\mathcal{O} 100\%$ and above. If this is true, it indicates that meaningful inferences can only be performed when we coherently combine information from a population of merger events, in a process known as coherent power-stacking. This is similar to Poisonian statistics and an error reduction by a factor of $1/\sqrt{N}$, where $N$ is the number of observations. Accordingly, we have performed a Monte-Carlo simulation of the comparable mass BBH merger events in an effort to make error estimates based on population-wide inferences.  However, in our case, it should be remembered that the error-reduction rate is not nearly as strong as $1/\sqrt{N}$, because all our events will not originate from the same distance. We skip the analogous exercise considered with EMRI systems, 
\subsubsection{Choice of populations} We have assumed events to be randomly and uniformly distributed per unit comoving volume element. For low redshifts $ z \lessapprox 0.5$, the rate of generation of BBH systems do not depend on the redshift $z$, and hence the above is a valid approximation to assume. We have also assumed events up to a maximum redshift of $z = 0.5$, after which it is expected that the sensitivities of current ground based detectors like aLIGO would decrease significantly. We consider two separate populations of 50000 and 100000 events respectively. Additionally for each of the mentioned cases, we assumed two kinds of mass-distributions in the populations. In the first case, we have assumed that the components of the binary are derived from a seed uniform distribution between $10M_\odot\leq m \leq 50 M_\odot$. The limits to the range are inspired from early models of supernova remnants. In the second case we have consider the other limit and assume that the component BHs are seeded from a relatively narrow  Gaussian distribution centered around $\mu = 50 M_\odot$ and a standard deviation of $\sigma = 5 M_\odot$.

\begin{figure}[h]
\centering
\includegraphics[width=8cm, height=6cm]{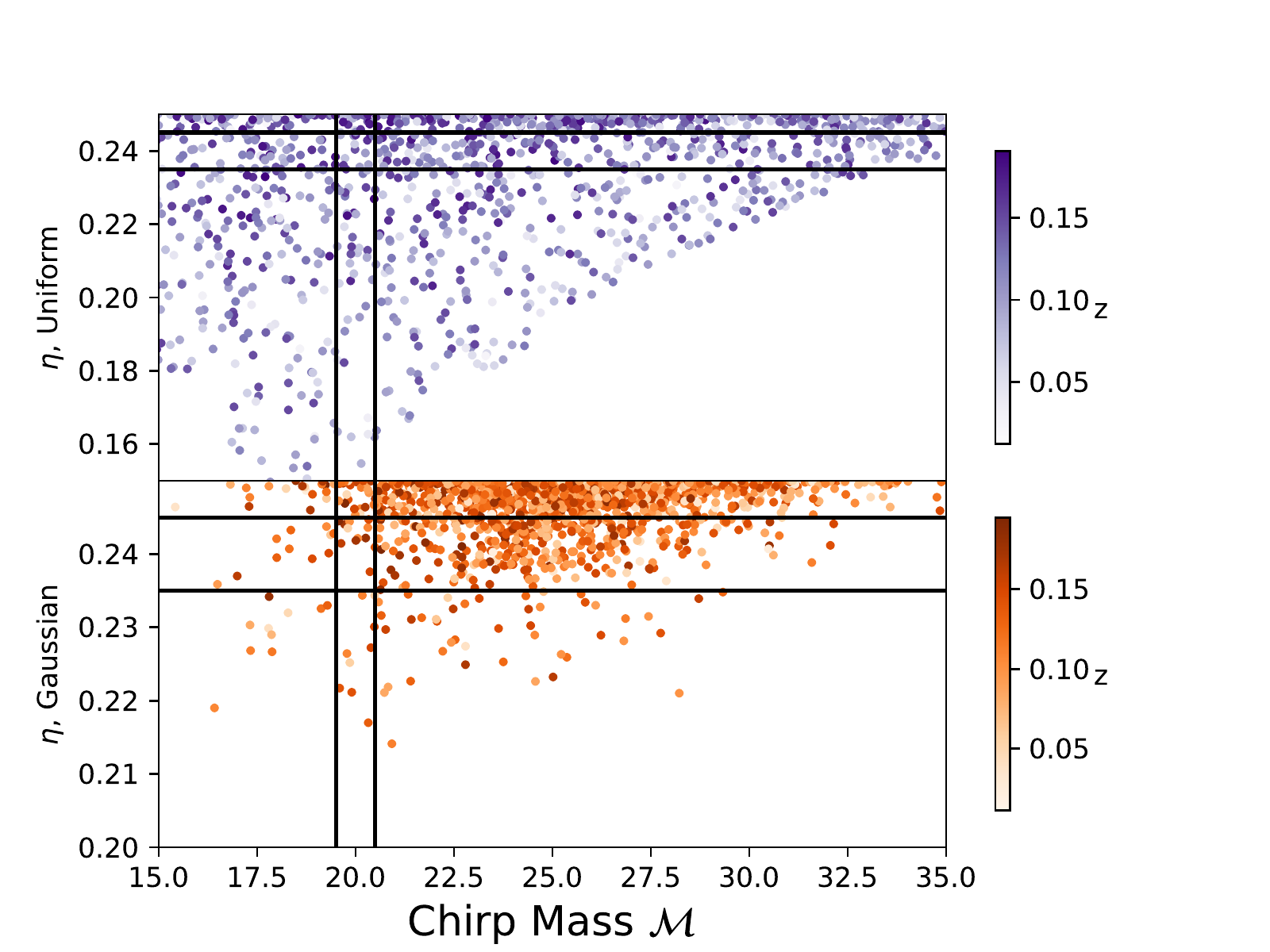}
\includegraphics[width=8cm, height=6cm]{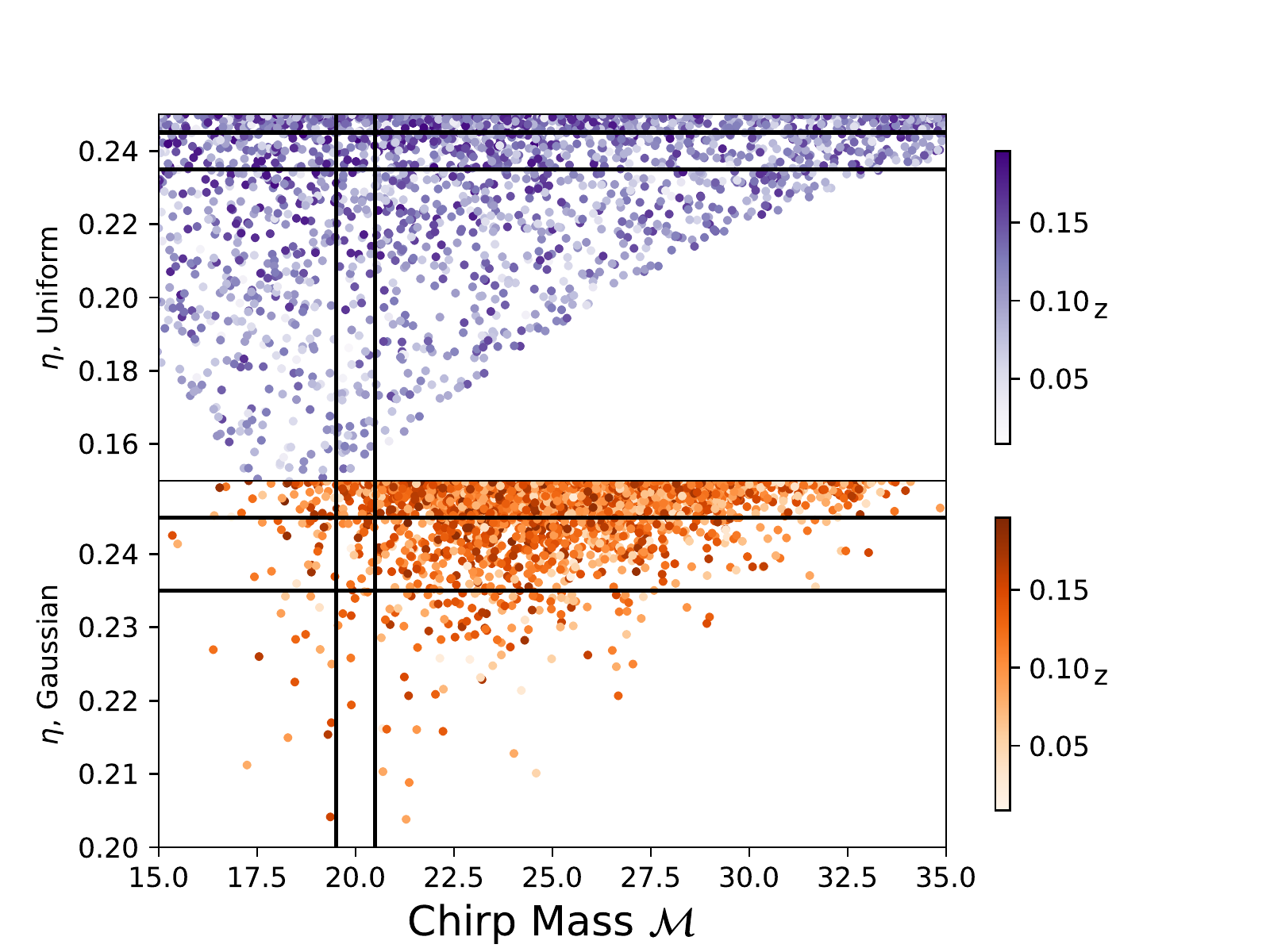}
\caption{Configuration of masses of populations along with redshift (shown as colour code) that have an SNR $\geq 20$ among 50000 events for the uniform mass distribution (upper panels) and the Gaussian mass distribution. The figure on the right shows exactly the same, but for 100000 events }
\label{fig: snr_threshold}
\end{figure} Additionally, we have assumed for every case that the binaries forming the population are uniformly distributed in their inclination $(0,\pi)$ and angle of polarisation $(0,2\pi)$. To set meaningful bounds on the initial eccentricity $e_0$, we first observe that as progenitors to systems of BBHs, given a separation binary main sequence stars have to have an upper limit in their eccentricity to avoid collision as they are extended bodies. Consequently the BBH system is expected to inherit an upper limit from it's progenitor binary main sequence star configuration. The details of this upper limit turn to be heavily model-dependent and are not central to our results. We have thus assumed initial eccentricity $0\leq e_0 \leq 0.5$. The choice of upper limit value to the eccentricity is ad-hoc. It should be also remarked here that this eccentricity is chosen when the orbital frequency $M\Omega \thicksim 10^{-4}$, and it is indeed so that all these binaries become almost circular when they are visible by ground-based detectors. Fig \ref{fig: snr_threshold} shows the members of the population that clear an SNR threshold of 20 for all cases. The black horizontal and vertical lines indicate a specific $\mathcal{M} = 20.0,\eta = 0.245$ bin, which is our bin of interest. Consequently, we combine information from all events that fall within this bin.

\subsubsection{Population-wide inference}
\begin{figure}[h]
\centering
\includegraphics[width=5cm, height=4cm]{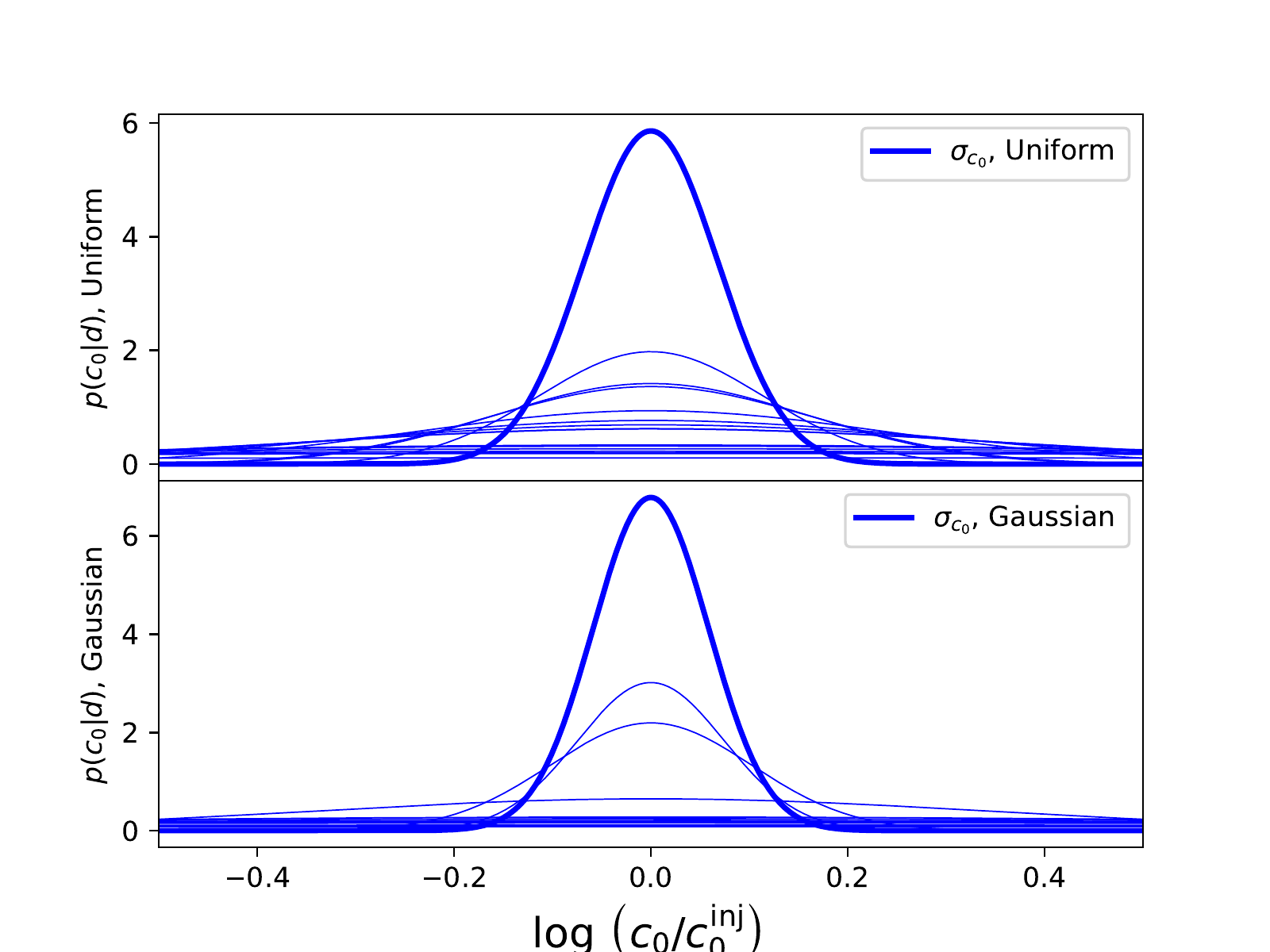}
\includegraphics[width=5cm, height=4cm]{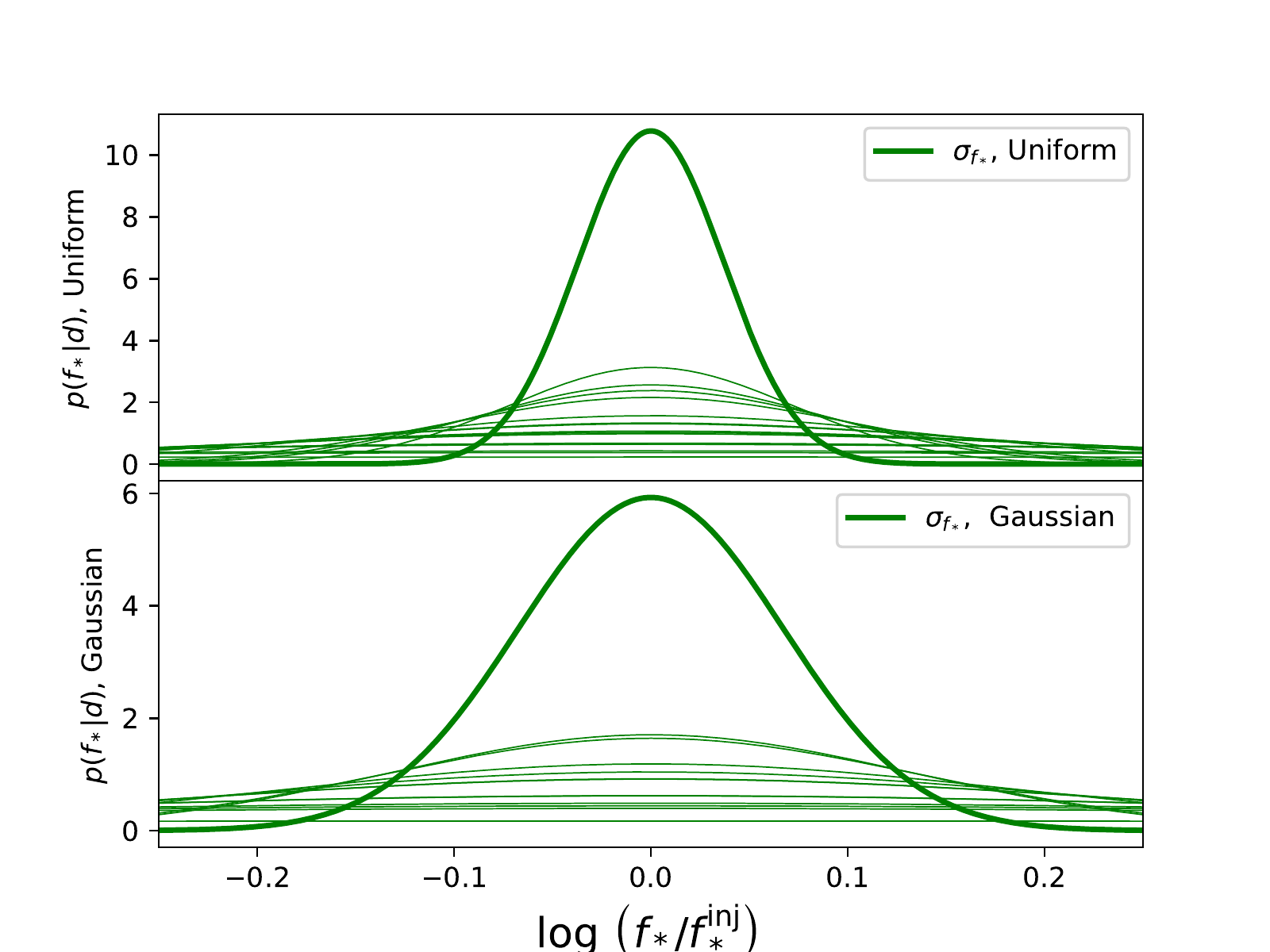}
\includegraphics[width=5cm, height=4cm]{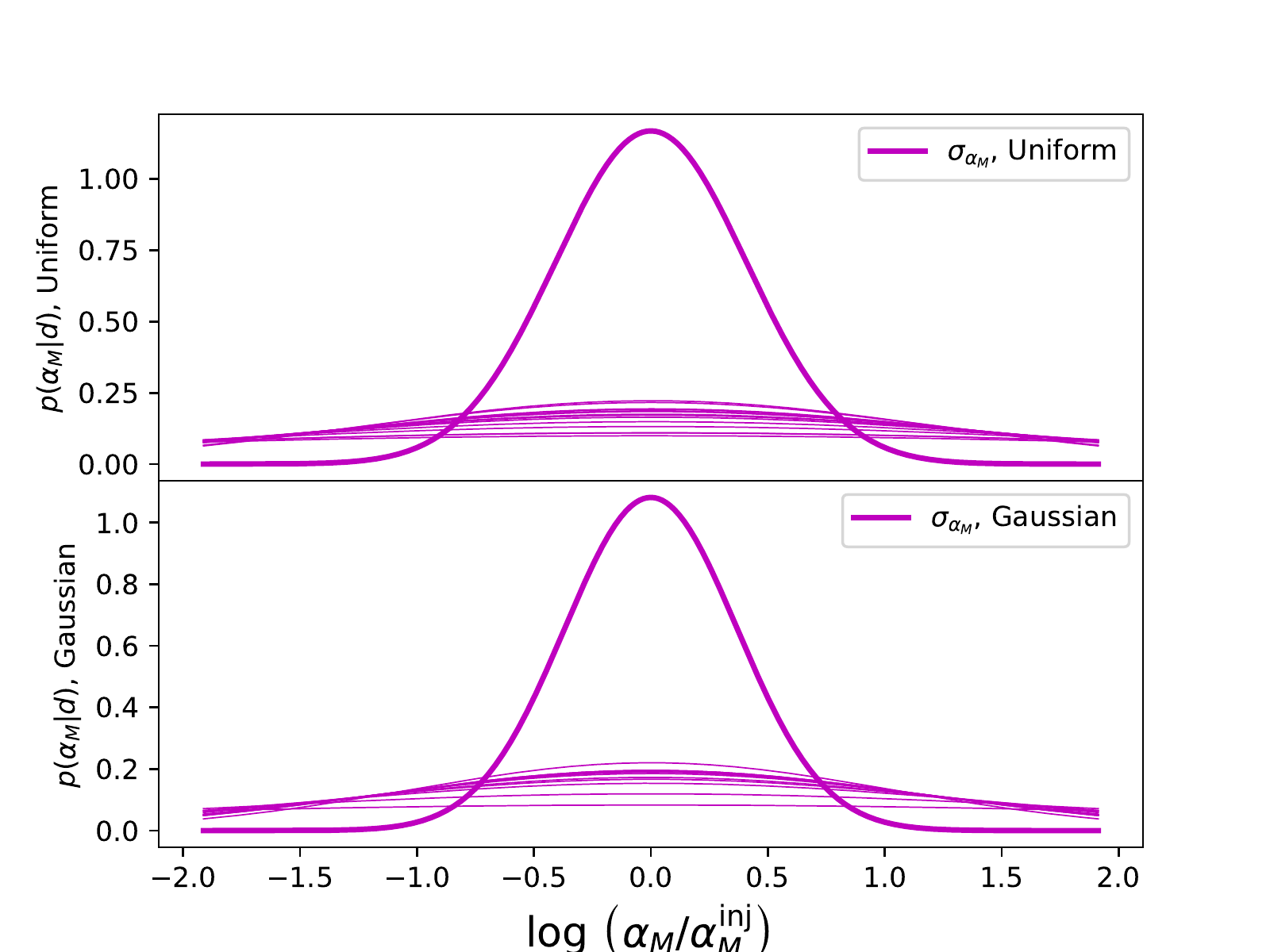} \\
\includegraphics[width=5cm, height=4cm]{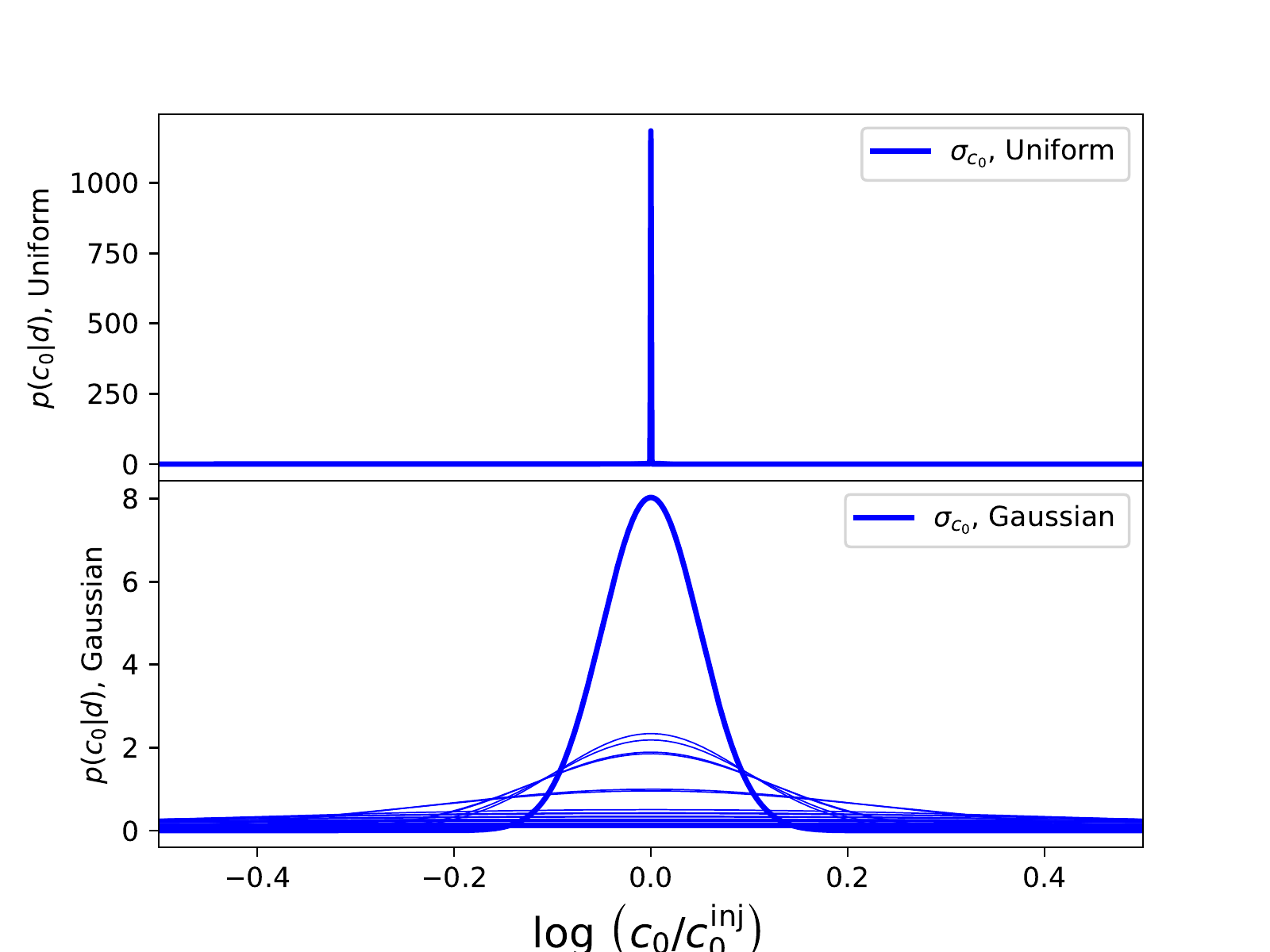}
\includegraphics[width=5cm, height=4cm]{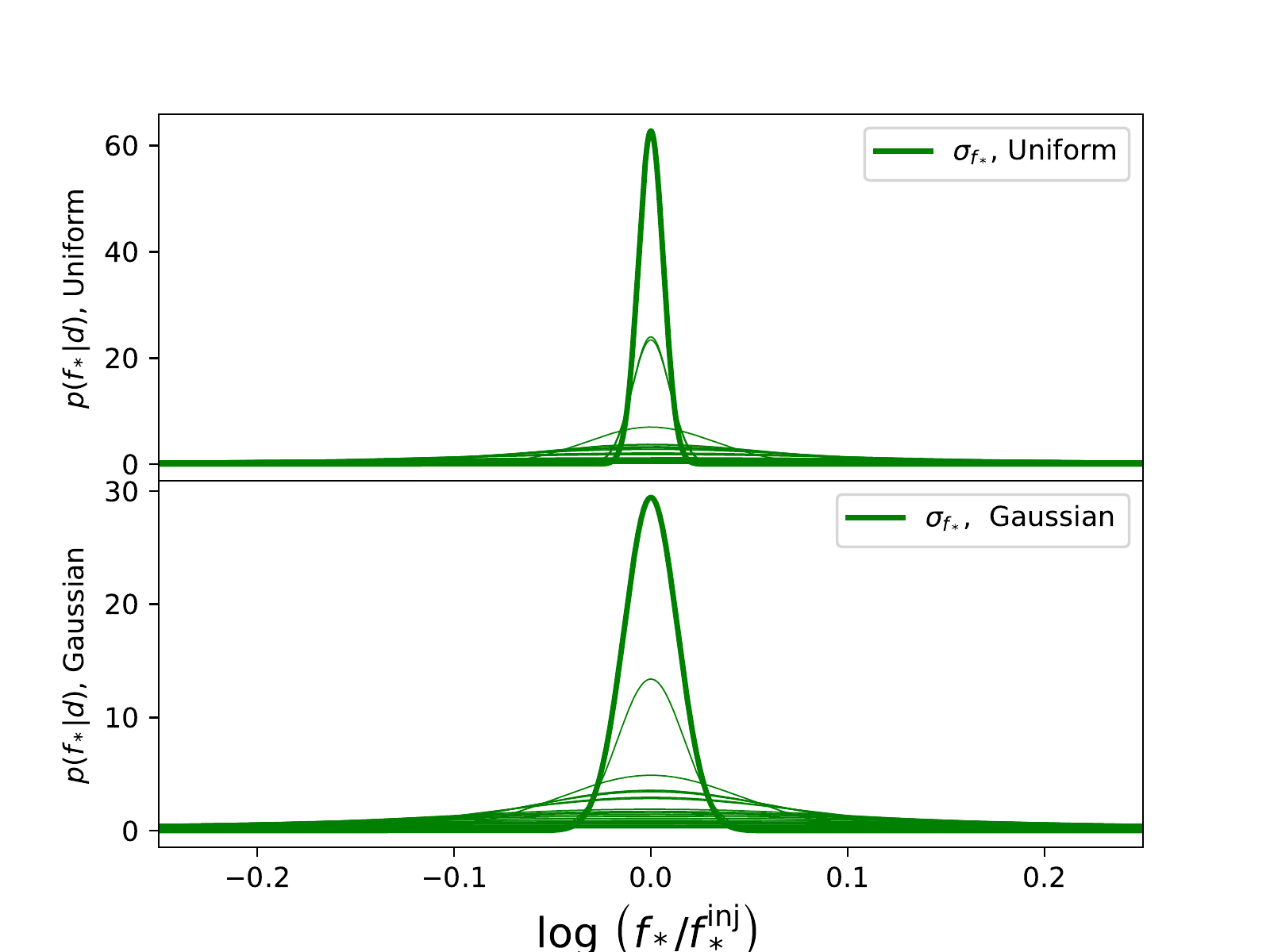}
\includegraphics[width=5cm, height=4cm]{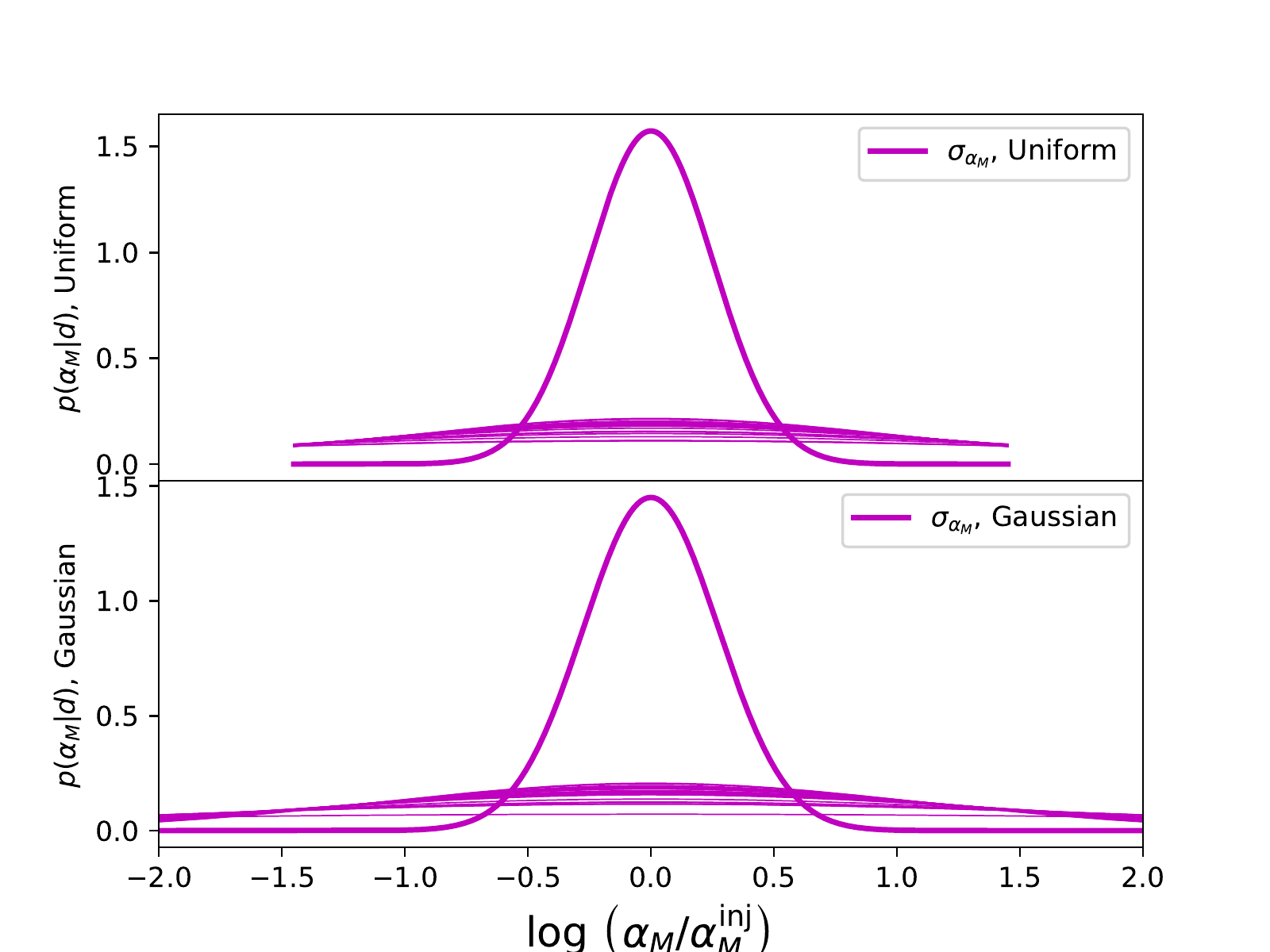}
\caption{Population-wide inference results for $c_0,f_*$ and $\alpha_M$ with the errors $\sigma_{c0}, \sigma_{f_*}$ and $\sigma_{\alpha_M}$ respectively for choice of $\mathcal{M} = 20$ and $\eta = 0.24$ with 50000 events (top panels) and 100000 events. The thin lines represent the errors of each individual event that crosses an SNR threshold of 20, while the thick lines represent the combined errors. The results are shown assuming the seed population of the components of the BBH come from uniform distribution (top sub-panels) and gaussian (bottom sub-panels) distributions.  panel  }
\label{fig:pop_infer}
\end{figure}

We are now in a position to analyse the results of our population-wide inferences. Fig \ref{fig:pop_infer} shows the results of the inference studies for $\alpha_T,\alpha_M$ for the binary populations seeded from uniform (upper sub-panels ) and gaussian (lower sub-panels) distributions of component mass. As stated before, the mass bin of our choice is $\mathcal{M} = 20.0,\eta = 0.245$.  The upper panel is the combined inference from 50000 events, while the lower one has 100000 events. We are drawn to make some important conclusions. First, we note that the errors in $\alpha_T$ which were $\thicksim 80-90 \%$ for $c_0$ and $\thicksim 10-15 \%$ for $f_*$ for a single event reduce to $\thicksim 10-20 \%$ and $\thicksim 5 \%$ respectively for the 50000 strong population survey. For the 100000 strong population, we find that the same errors on $\alpha_T$ get constrained to less than $5\%$ for $c_0$ and less than $1.3\%$ for $f_*$. It is thus clear that EMRIs are the cleanest probes of the $\alpha_T$ subspace. However, until LISA comes online EMRIS are not possible to observe, and population-wide inferences of $\alpha_T$ turn out to be useful tools in constraining them. We further note that in order to achieve constraints on  $\alpha_T$ to a few percent, we realistically need a population of 100000 strong events which translates to a total of $\thicksim 4000$ detections. With a detection rate of $\thicksim 10$ per year we can see that the current generation of ground based GW detetors will most likely not be able to resolve these effects. However, the 3G network of ground based GW detectors scheduled around 2030s are expected to detect tens of thousands of event per year, and would thus be able to resolve $\alpha_T$ to a few percent with $\thicksim 1$ year of data. We thus demonstrate that populations of comparable mass inspirals can produce meaningful constraints with just 1 year of 3G data and can potentially narrow the $\alpha_T$ parameter space for EMRI inference.

\paragraph*{} Single event EMRIs, like their comparable mass inspiral counterparts will nevertheless fail to constrain the $\alpha_M$ subspace. As the formation channels of supermassive BHs are not clearly understood, we are handicapped in modelling realistic populations of EMRIs. In this case therefore, the population-wide inference of comparable mass binaries happens to be our only option. As the single event inferences produce error-margins which are $\thicksim 300-400\%$ it is much harder to constrain $\alpha_M$. We show the results of population-wide $\alpha_M$ inference in the third coloumn of Fig \ref{fig:pop_infer}. We compute that with the 100000 strong population ( or $\thicksim 4000$ detections) $\alpha_M$ is constrained to an accuracy of nearly $25\%$.

\paragraph*{}Estimates from population-wide inferences are expected to be dependent upon the nature of the populations themselves. But how big is the dependence, and at what threshold does it begin to appear ? The first question was exactly the premise of the investigations done in \cite{Leyde:2022fsc}. We want to see if these questions show up in our results as well. We find from Fig \ref{fig:pop_infer} that interestingly, the $\alpha_M$ inferences are comparatively less sensitive to differences in population, as compared to the $\alpha_T$ subspace. This observation can once again be explained by considering that variation in component mass over populations have much less effect on the amplitude of GWs, as compared to their phase. Focussing on the $\alpha_T = (c_), f_*)$ subspace, our results also show that differences across seed populations show up more starkly (particularly in $c_0$) with 100000 events (4000 detections) . Hence we expect this number of detections to be the threshold of such effects of differences in population to show up.

\paragraph*{} The narrow posteriors of $c_0$ for both sets of populations lead us to conclude that as a parameter, it will be statistically distinguishable from the baseline model which in our case is just $\Lambda$CDM. We also note that for $\Lambda$CDM $f_*$ is undefined, as will be the case if $c_0 = 0$ in \ref{EFT_ansatz}. However, the same cannot be directly said for the parameter $\alpha_M$, because of the presence of tens of percent of error even with the 100000 strong population. To quantify this uncertainty, we have run an equivalent inference on $c_0$ and $\alpha_M$ for the baseline model $\Lambda$CDM. shows us the results. As expected, for both sets of populations $c_0$ turns out to be statistically distinguishable, given our choice of $c_0 = 0.2$. With $\alpha_M$, we see that for the 100000 strong population for both the gaussian and uniform distributions, the posterior for $\alpha_M = 0.7$ intersects the corresponding one for $\alpha_M= 0$ just over $2 \sigma$. We are thus led to believe that a $2\sigma$ significant detection over $\Lambda$CDM is probable for $\alpha_M$ with $\thicksim 4000$ detections.    

\section{Conclusions}
GWs offer a reliable window into understanding if beyond-$\Lambda$CDM models of gravity could be meaningfully inferred. From this work, we are able to arrive at several independent and important conclusions. We have performed inferences of the beyond-$\Lambda$CDM parameters $\alpha_T,\alpha_M$ with single events as well as with population-wide surveys. Our Fisher estimates for single events shows correlations in between the parameters which depend on eccentricity which demonstrates how unmodelled eccentricity can silently bias inference studies. We also find that EMRIs by virtue of their long inspiral times have the best chances of inferring the $\alpha_T$ subspace just in a single event. However, owing to the presence of seismic cut-off this task cannot be performed by ground-based detectors. Space based missions like eLISA can be our answer here. Furthermore we show that even EMRI syatems cannot resolve $\alpha_M$ by a single event. In order to get around this problem, we successfully demonstrate the power of population-wide inferences of comparable mass binary merger systems as a tool to constrain the otherwise poorly constrained $\alpha_M$ to $\thicksim25\%$. With such a constraint, we can infer $\alpha_M$ over its baseline value of 0 to $2\sigma \approx 95\%$ confidence interval. It turns out that this is the best we can do with $\alpha_M$. However, given our state-of-art of GW detectors, we cannot achieve this target in a reasonable amount of time as the most accurate inference will take $\thicksim 4000$ detections on the average. Fortunately, the number of detections necessary are right in the ballpark of the upcomiong 3G detector network thanks to their enhanced detection rates. With 3G detectors, $\thicksim 4000$ detections would take around 1 year, which would be the time-frame necessary to acheive the accuracy we calculate.

\begin{figure}[h]
\centering
\includegraphics[width=12cm, height=9cm]{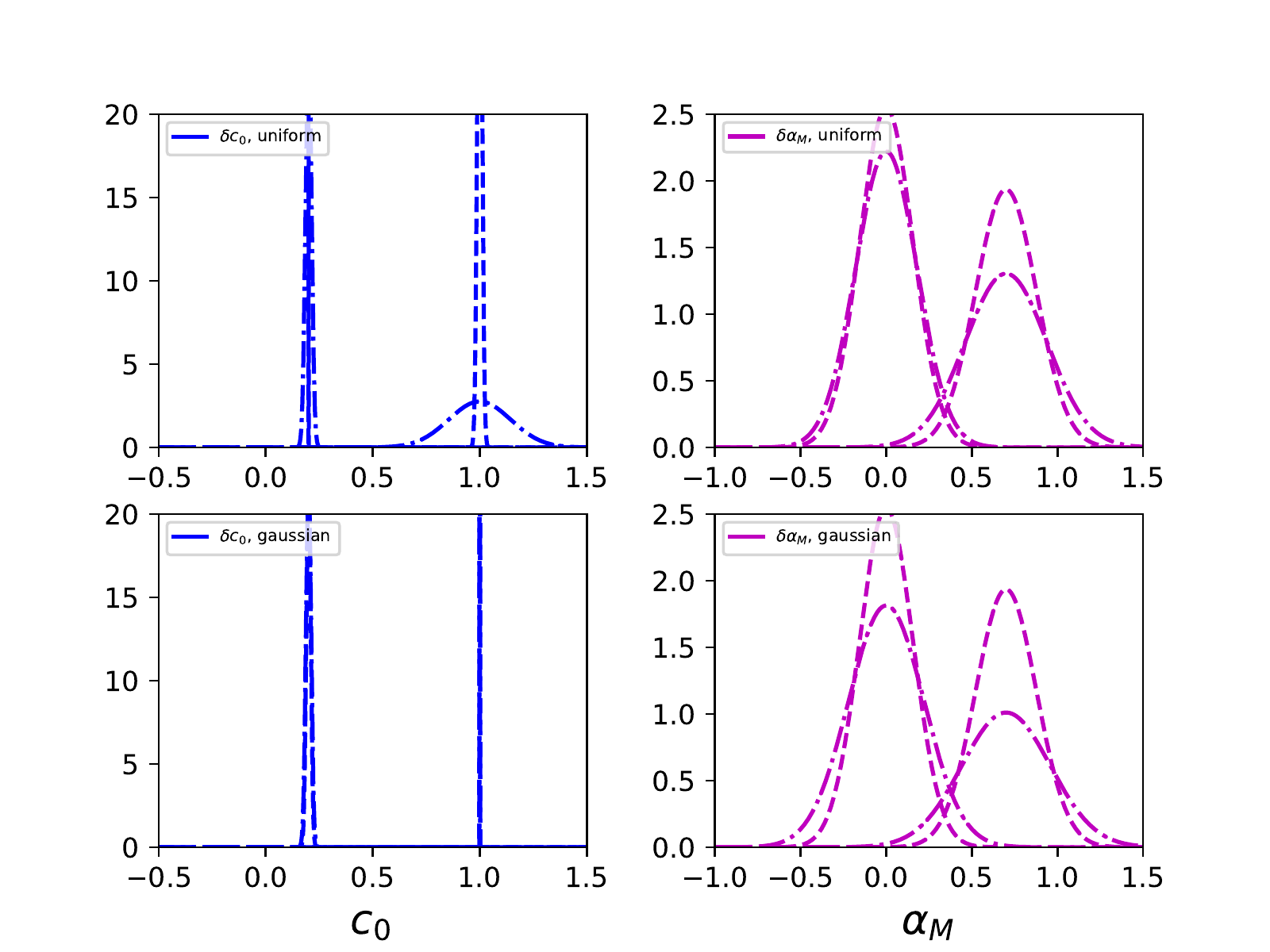}
\caption{Comparison of our choice of inference of beyond-$\Lambda$CDM parameters $c_0 = 0.2$ and $\alpha_M = 0.7$ with the corresponding baseline  $\Lambda$CDM model, which has $c_0 = 1.0$ and $\alpha_M = 0$. Top panels show the uniform populations, the bottom ones show the gaussian populations. While $c_0$ is easily distinguishable, $\alpha_M$ can be distinguished only to $2\sigma$ accuracy with the 100000 strong population}
\label{fig: baseline}
\end{figure}

\paragraph*{} For a single event, we have also considered the effects of modelling inaccuracies at source, by progressively decreasing the PN accuracy order-by-order. We demonstrate that in a counter-intuitive twist, lowering of the order of PN accuracy does affect the outcomes of inference study of propagation parameters, particularly when dealing with eccentric binaries. We anticipate that such kind of modelling runs the risk of being a nuisance by giving rise to a large source of bias in the $\alpha_T,\alpha_M$. In order to mitigate this problem, one must therefore perform a more complete and accurate source modelling, namely including higher order PN eccentric tail and tail-of-tail dependent terms. In addition, the inclusion of spin dependent terms like individual spins, spin-orbit and spin-spin couplings are expected to introduce precession of the binary orbits, which will also modify the beyond-$\Lambda$CDM inference estimates. Refinements would also need to include the current state-of-art at 4 PN order.

\paragraph*{} In our present work, we have attempted a simplistic Fisher forecast study for beyond $\Lambda$CDM parameters with three situations, namely inference single events (EMRIs or comparable mass inspirals), studying the effects of PN orders on inferences, and considering population-wide inferences. Our choice of binary configurations was also simplified by ignoring spin and higher order source effects. A subsequent study would thus have to include these effects at source and perform a full Bayesian sampling of the multidimensional likelihood function for the beyond $\Lambda$CDM subspace. We leave such an exercise for a future attempt.

\section{Acknowledgements}

This work is a result of early discussions with Ippocratis Saltas and Roberto Oliveri. It has partly been supported by the grant from the Czech Academy of Sciences under Project No. LQ10010210. K.C acknowledges additional helpful inputs from Ignacy Sawicki, Luc Blanchet, David Trestini, and Georgois Loukes Gerakoupoulos. K.C also acknowledges the use of the Phoebe cluster at CEICO, FZU and I.T. support by Josef Dvoracek.

\bibliographystyle{unsrt}
\bibliography{ref}

\end{document}